\documentclass[twocolumn,twocolappendix]{aastex631}

\usepackage{graphicx}
\usepackage{amsmath}
\usepackage{multirow}
\usepackage{todonotes}
\usepackage{float}
\usepackage{lipsum}
\usepackage{soul}

\newcommand{\rev}[1]{{#1}} %veritas-related revisions
\newcommand{\frev}[1]{{#1}} %fermi-related revisions
\newcommand{\hrev}[1]{{#1}} %hawc-related revisions
\newcommand{\apjrev}[1]{{{#1}}}
\begin{document}

\title{A Multiwavelength Interpretation of HESS~J1857+026 Emission Using the Fermi--LAT, VERITAS, and HAWC Observatories}

\collaboration{2}{The Fermi-LAT collaboration}

\author[0000-0001-9633-3165]{J. Eagle}
\affiliation{
NASA Goddard Space Flight Center,
Greenbelt, MD, 20771, USA}

\author[0000-0002-5167-1221]{S. Kumar}
\affiliation{
Department of Physics and Astronomy, University of Maryland, 
College Park, Maryland 20742, USA}

\collaboration{100}{The VERITAS collaboration}

\author{A.~Archer}\affiliation{Department of Physics and Astronomy, DePauw University, Greencastle, IN 46135-0037, USA}
\author[0000-0002-3886-3739]{P.~Bangale}\affiliation{Department of Physics, Temple University, Philadelphia, PA 19122, USA}
\author[0000-0002-9675-7328]{J.~T.~Bartkoske}\affiliation{Department of Physics and Astronomy, University of Utah, Salt Lake City, UT 84112, USA}
\author[0000-0003-2098-170X]{W.~Benbow}\affiliation{Center for Astrophysics $|$ Harvard \& Smithsonian, Cambridge, MA 02138, USA}
\author{N.~R.~Bond}\affiliation{School of Physics, University College Dublin, Belfield, Dublin 4, Ireland}
\author[0009-0001-5719-936X]{Y.~Chen}\affiliation{Department of Physics and Astronomy, University of California, Los Angeles, CA 90095, USA}
\author[0000-0001-5811-9678]{J.~L.~Christiansen}\affiliation{Physics Department, California Polytechnic State University, San Luis Obispo, CA 94307, USA}
\author{A.~J.~Chromey}\affiliation{Center for Astrophysics $|$ Harvard \& Smithsonian, Cambridge, MA 02138, USA}
\author[0000-0003-1716-4119]{A.~Duerr}\affiliation{Department of Physics and Astronomy, University of Utah, Salt Lake City, UT 84112, USA}
\author[0000-0002-1853-863X]{M.~Errando}\affiliation{Department of Physics, Washington University, St. Louis, MO 63130, USA}
\author{M.~Escobar~Godoy}\affiliation{Santa Cruz Institute for Particle Physics and Department of Physics, University of California, Santa Cruz, CA 95064, USA}
\author[0000-0002-4131-655X]{J.~Escudero~Pedrosa}\affiliation{Center for Astrophysics $|$ Harvard \& Smithsonian, Cambridge, MA 02138, USA}
\author{S.~Feldman}\affiliation{Department of Physics and Astronomy, University of California, Los Angeles, CA 90095, USA}
\author[0000-0001-6674-4238]{Q.~Feng}\affiliation{Department of Physics and Astronomy, University of Utah, Salt Lake City, UT 84112, USA}
\author[0000-0002-2636-4756]{S.~Filbert}\affiliation{Department of Physics and Astronomy, University of Utah, Salt Lake City, UT 84112, USA}
\author[0000-0002-1067-8558]{L.~Fortson}\affiliation{School of Physics and Astronomy, University of Minnesota, Minneapolis, MN 55455, USA}
\author[0000-0003-1614-1273]{A.~Furniss}\affiliation{Santa Cruz Institute for Particle Physics and Department of Physics, University of California, Santa Cruz, CA 95064, USA}
\author[0000-0002-0109-4737]{W.~Hanlon}\affiliation{Center for Astrophysics $|$ Harvard \& Smithsonian, Cambridge, MA 02138, USA}
\author[0000-0001-6951-2299]{C.~E.~Hinrichs}\affiliation{Center for Astrophysics $|$ Harvard \& Smithsonian, Cambridge, MA 02138, USA and Department of Physics and Astronomy, Dartmouth College, 6127 Wilder Laboratory, Hanover, NH 03755 USA}
\author[0000-0002-6833-0474]{J.~Holder}\affiliation{Department of Physics and Astronomy and the Bartol Research Institute, University of Delaware, Newark, DE 19716, USA}
\author[0000-0002-1432-7771]{T.~B.~Humensky}\affiliation{Department of Physics, University of Maryland, College Park, MD, USA and NASA GSFC, Greenbelt, MD 20771, USA}
\author{M.~Iskakova}\affiliation{Department of Physics, Washington University, St. Louis, MO 63130, USA}
\author[0000-0002-1089-1754]{W.~Jin}\affiliation{Department of Physics and Astronomy, University of California, Los Angeles, CA 90095, USA}
\author[0009-0008-2688-0815]{M.~N.~Johnson}\affiliation{Santa Cruz Institute for Particle Physics and Department of Physics, University of California, Santa Cruz, CA 95064, USA}
\author[0000-0001-8557-1141]{E.~Joshi}\affiliation{DESY, Platanenallee 6, 15738 Zeuthen, Germany}
\author[0000-0002-3638-0637]{P.~Kaaret}\affiliation{Department of Physics and Astronomy, University of Iowa, Van Allen Hall, Iowa City, IA 52242, USA}
\author{M.~Kertzman}\affiliation{Department of Physics and Astronomy, DePauw University, Greencastle, IN 46135-0037, USA}
\author{M.~Kherlakian}\affiliation{Fakult\"at f\"ur Physik \& Astronomie, Ruhr-Universit\"at Bochum, D-44780 Bochum, Germany}
\author[0000-0003-4785-0101]{D.~Kieda}\affiliation{Department of Physics and Astronomy, University of Utah, Salt Lake City, UT 84112, USA}
\author[0000-0002-4260-9186]{T.~K.~Kleiner}\affiliation{DESY, Platanenallee 6, 15738 Zeuthen, Germany}
\author[0000-0002-4289-7106]{N.~Korzoun}\affiliation{Department of Physics and Astronomy and the Bartol Research Institute, University of Delaware, Newark, DE 19716, USA}
\author[0000-0002-5167-1221]{S.~Kumar}\affiliation{Department of Physics, University of Maryland, College Park, MD, USA }
\author{S.~Kundu}\affiliation{Department of Physics and Astronomy, University of Alabama, Tuscaloosa, AL 35487, USA}
\author[0000-0003-3802-1619]{M.~Lundy}\affiliation{Physics Department, McGill University, Montreal, QC H3A 2T8, Canada}
\author[0000-0001-9868-4700]{G.~Maier}\affiliation{DESY, Platanenallee 6, 15738 Zeuthen, Germany}
\author[0000-0001-7106-8502]{M.~J.~Millard}\affiliation{Department of Physics and Astronomy, University of Iowa, Van Allen Hall, Iowa City, IA 52242, USA}
\author[0000-0002-1499-2667]{P.~Moriarty}\affiliation{School of Natural Sciences, University of Galway, University Road, Galway, H91 TK33, Ireland}
\author[0000-0002-3223-0754]{R.~Mukherjee}\affiliation{Department of Physics and Astronomy, Barnard College, Columbia University, NY 10027, USA}
\author[0000-0002-4837-5253]{R.~A.~Ong}\affiliation{Department of Physics and Astronomy, University of California, Los Angeles, CA 90095, USA}
\author[0000-0003-3820-0887]{A.~Pandey}\affiliation{Department of Physics and Astronomy, University of Utah, Salt Lake City, UT 84112, USA}
\author[0000-0001-7861-1707]{M.~Pohl}\affiliation{Institute of Physics and Astronomy, University of Potsdam, 14476 Potsdam-Golm, Germany and DESY, Platanenallee 6, 15738 Zeuthen, Germany}
\author[0000-0002-0529-1973]{E.~Pueschel}\affiliation{Fakult\"at f\"ur Physik \& Astronomie, Ruhr-Universit\"at Bochum, D-44780 Bochum, Germany}
\author[0000-0002-5104-5263]{P.~L.~Rabinowitz}\affiliation{Department of Physics, Washington University, St. Louis, MO 63130, USA}
\author[0000-0002-5351-3323]{K.~Ragan}\affiliation{Physics Department, McGill University, Montreal, QC H3A 2T8, Canada}
\author{P.~T.~Reynolds}\affiliation{Department of Physical Sciences, Munster Technological University, Bishopstown, Cork, T12 P928, Ireland}
\author{E.~Roache}\affiliation{Center for Astrophysics $|$ Harvard \& Smithsonian, Cambridge, MA 02138, USA}
\author[0000-0003-1387-8915]{I.~Sadeh}\affiliation{DESY, Platanenallee 6, 15738 Zeuthen, Germany}
\author[0000-0002-3171-5039]{L.~Saha}\affiliation{Center for Astrophysics $|$ Harvard \& Smithsonian, Cambridge, MA 02138, USA}
\author[0009-0000-0295-8800]{H.~Salzmann}\affiliation{Santa Cruz Institute for Particle Physics and Department of Physics, University of California, Santa Cruz, CA 95064, USA}
\author{M.~Santander}\affiliation{Department of Physics and Astronomy, University of Alabama, Tuscaloosa, AL 35487, USA}
\author{G.~H.~Sembroski}\affiliation{Department of Physics and Astronomy, Purdue University, West Lafayette, IN 47907, USA}
\author[0000-0002-9856-989X]{R.~Shang}\affiliation{Department of Physics and Astronomy, Barnard College, Columbia University, NY 10027, USA}
\author[0009-0008-7331-7240]{S.~Tandon}\affiliation{Physics Department, Columbia University, New York, NY 10027, USA}
\author{J.~V.~Tucci}\affiliation{Department of Physics, Indiana University Indianapolis, Indianapolis, Indiana 46202, USA}
\author{V.~V.~Vassiliev}\affiliation{Department of Physics and Astronomy, University of California, Los Angeles, CA 90095, USA}
\author[0000-0003-2740-9714]{D.~A.~Williams}\affiliation{Santa Cruz Institute for Particle Physics and Department of Physics, University of California, Santa Cruz, CA 95064, USA}
\author[0000-0002-2730-2733]{S.~L.~Wong}\affiliation{Physics Department, McGill University, Montreal, QC H3A 2T8, Canada}
\author{T.~Yoshikoshi}\affiliation{Institute for Cosmic Ray Research, University of Tokyo, 5-1-5, Kashiwa-no-ha, Kashiwa, Chiba 277-8582, Japan}
% \author[0000-0002-9856-989X]{Ruo-Yu Shang}
% \affiliation{
% Department of Physics and Astronomy, Barnard College, Columbia University,
% New York, NY 10027, USA}

% \author[0009-0001-5719-936X]{Yu Chen}
% \affiliation{Department of Physics and Astronomy, University of California, Los Angeles, CA 90095, USA} 

\collaboration{100}{The HAWC collaboration}

\author[0000-0002-7747-754X]{S.~Coutiño de León}
\affiliation{Department of Physics, University of Wisconsin-Madison, Madison, WI, USA} 

\author[0000-0002-7102-3352]{R. Torres-Escobedo}
\affiliation{Tsung-Dao Lee Institute \& School of Physics and Astronomy, Shanghai Jiao Tong University, Shanghai 201210, China}

% HAWC author list
\author{R.~Alfaro}
\affiliation{Instituto de F\'{i}sica, Universidad Nacional Autónoma de México, Ciudad de Mexico, Mexico }

\author{C.~Alvarez}
\affiliation{Universidad Autónoma de Chiapas, Tuxtla Gutiérrez, Chiapas, México}

\author{E.~Anita-Rangel}
\affiliation{Instituto de Astronom\'{i}a, Universidad Nacional Autónoma de México, Ciudad de Mexico, Mexico}

\author{M.~Araya}
\affiliation{Universidad de Costa Rica, San José 2060, Costa Rica}

\author{J.C.~Arteaga-Velázquez}
\affiliation{Universidad Michoacana de San Nicolás de Hidalgo, Morelia, Mexico }

\author{D.~Avila Rojas}
\affiliation{Instituto de Astronom\'{i}a, Universidad Nacional Autónoma de México, Ciudad de Mexico, Mexico}

\author{R.~Babu}
\affiliation{Department of Physics and Astronomy, Michigan State University, East Lansing, MI, USA}

\author{P.~Bangale}
\affiliation{Temple University, Department of Physics, 1925 N. 12th Street, Philadelphia, PA 19122, USA}

\author{E.~Belmont-Moreno}
\affiliation{Instituto de F\'{i}sica, Universidad Nacional Autónoma de México, Ciudad de Mexico, Mexico}

\author{A.~Bernal}
\affiliation{Instituto de Astronom\'{i}a, Universidad Nacional Autónoma de México, Ciudad de Mexico, Mexico}

\author{K.S.~Caballero-Mora}
\affiliation{Universidad Autónoma de Chiapas, Tuxtla Gutiérrez, Chiapas, México}

\author{T.~Capistrán}
\affiliation{Università degli Studi di Torino, I-10125 Torino, Italy}

\author{F.~Carreón}
\affiliation{Instituto de Astronom\'{i}a, Universidad Nacional Autónoma de México, Ciudad de Mexico, Mexico}

\author{S.~Casanova}
\affiliation{Institute of Nuclear Physics Polish Academy of Sciences, PL-31342 IFJ-PAN, Krakow, Poland}

\author{U.~Cotti}
\affiliation{Universidad Michoacana de San Nicolás de Hidalgo, Morelia, Mexico}

\author{J.~Cotzomi}
\affiliation{Facultad de Ciencias F\'{i}sico Matemáticas, Benemérita Universidad Autónoma de Puebla, Puebla, Mexico}

\author{E.~De la Fuente}
\affiliation{Departamento de F\'{i}sica, Centro Universitario de Ciencias Exactase Ingenierias, Universidad de Guadalajara, Guadalajara, Mexico}

\author{P.~Desiati}
\affiliation{Dept. of Physics and Wisconsin IceCube Particle Astrophysics Center, University of Wisconsin{\textemdash}Madison, Madison, WI, USA}

\author{N.~Di Lalla}
\affiliation{Department of Physics, Stanford University: Stanford, CA 94305–4060, USA}

\author{R.~Diaz Hernandez}
\affiliation{Instituto Nacional de Astrof\'{i}sica, Óptica y Electrónica, Puebla, Mexico}

\author{B.L.~Dingus}
\affiliation{Los Alamos National Laboratory, Los Alamos, NM, USA}

\author{M.A.~DuVernois}
\affiliation{Dept. of Physics and Wisconsin IceCube Particle Astrophysics Center, University of Wisconsin{\textemdash}Madison, Madison, WI, USA}

\author{J.C.~Díaz-Vélez}
\affiliation{Dept. of Physics and Wisconsin IceCube Particle Astrophysics Center, University of Wisconsin{\textemdash}Madison, Madison, WI, USA}

\author{K.~Engel}
\affiliation{Department of Physics, University of Maryland, College Park, MD, USA}

\author{T.~Ergin}
\affiliation{Department of Physics and Astronomy, Michigan State University, East Lansing, MI, USA}

\author{C.~Espinoza}
\affiliation{Instituto de F\'{i}sica, Universidad Nacional Autónoma de México, Ciudad de Mexico, Mexico}

\author{N.~Fraija}
\affiliation{Instituto de Astronom\'{i}a, Universidad Nacional Autónoma de México, Ciudad de Mexico, Mexico}

\author{J.A.~García-González}
\affiliation{Tecnologico de Monterrey, Escuela de Ingenier\'{i}a y Ciencias, Ave. Eugenio Garza Sada 2501, Monterrey, N.L., Mexico, 64849}

\author{F.~Garfias}
\affiliation{Instituto de Astronom\'{i}a, Universidad Nacional Autónoma de México, Ciudad de Mexico, Mexico}

\author{N.~Ghosh}
\affiliation{Department of Physics and Astronomy, Michigan State University, East Lansing, MI, USA}

\author{M.M.~González}
\affiliation{Instituto de Astronom\'{i}a, Universidad Nacional Autónoma de México, Ciudad de Mexico, Mexico}

\author{J.A.~González}
\affiliation{Universidad Michoacana de San Nicolás de Hidalgo, Morelia, Mexico}

\author{J.~Gyeong}
\affiliation{Department of Physics, Sungkyunkwan University, Suwon 16419, South Korea}

\author{J.P.~Harding}
\affiliation{Los Alamos National Laboratory, Los Alamos, NM, USA}

\author{S.~Hernández-Cadena}
\affiliation{Tsung-Dao Lee Institute \& School of Physics and Astronomy, Shanghai Jiao Tong University, 800 Dongchuan Rd, Shanghai, SH 200240, China}

\author{I.~Herzog}
\affiliation{Department of Physics and Astronomy, Michigan State University, East Lansing, MI, USA}

\author{D.~Huang}
\affiliation{University of Delaware, Department of Physics and Astronomy, Newark, DE, USA}

\author{F.~Hueyotl-Zahuantitla}
\affiliation{Universidad Autónoma de Chiapas, Tuxtla Gutiérrez, Chiapas, México}

\author{A.~Iriarte}
\affiliation{Instituto de Astronom\'{i}a, Universidad Nacional Autónoma de México, Ciudad de Mexico, Mexico}

\author{S.~Kaufmann}
\affiliation{Universidad Politecnica de Pachuca, Pachuca, Hgo, Mexico}

\author{J.~Lee}
\affiliation{University of Seoul, Seoul, Rep. of Korea}

\author{H.~León Vargas}
\affiliation{Instituto de Astronom\'{i}a, Universidad Nacional Autónoma de México, Ciudad de Mexico, Mexico}

\author{A.L.~Longinotti}
\affiliation{Instituto de Astronom\'{i}a, Universidad Nacional Autónoma de México, Ciudad de Mexico, Mexico}

\author{G.~Luis-Raya}
\affiliation{Universidad Politecnica de Pachuca, Pachuca, Hgo, Mexico}

\author{K.~Malone}
\affiliation{Los Alamos National Laboratory, Los Alamos, NM, USA}

\author{J.~Martínez-Castro}
\affiliation{Centro de Investigaci\'on en Computaci\'on, Instituto Polit\'ecnico Nacional, M\'exico City, M\'exico.}

\author{J.A.~Matthews}
\affiliation{Dept of Physics and Astronomy, University of New Mexico, Albuquerque, NM, USA}

\author{P.~Miranda-Romagnoli}
\affiliation{Universidad Autónoma del Estado de Hidalgo, Pachuca, Mexico}

\author{M.~Mostafá}
\affiliation{Temple University, Department of Physics, 1925 N. 12th Street, Philadelphia, PA 19122, USA}

\author{H.A.~Ayala Solares}
\affiliation{Temple University, Department of Physics, 1925 N. 12th Street, Philadelphia, PA 19122, USA}

\author{M.~Najafi}
\affiliation{Department of Physics, Michigan Technological University, Houghton, MI, USA}

\author{A.~Nayerhoda}
\affiliation{Institute of Nuclear Physics Polish Academy of Sciences, PL-31342 IFJ-PAN, Krakow, Poland}

\author{L.~Nellen}
\affiliation{Instituto de Ciencias Nucleares, Universidad Nacional Autónoma de Mexico, Ciudad de Mexico, Mexico}

\author{N.~Omodei}
\affiliation{Department of Physics, Stanford University: Stanford, CA 94305–4060, USA}

\author{E.~Ponce}
\affiliation{Facultad de Ciencias F\'{i}sico Matemáticas, Benemérita Universidad Autónoma de Puebla, Puebla, Mexico}

\author{E.G.~Pérez-Pérez}
\affiliation{Universidad Politecnica de Pachuca, Pachuca, Hgo, Mexico}

\author{C.D.~Rho}
\affiliation{Department of Physics, Sungkyunkwan University, Suwon 16419, South Korea}

\author{A.~Rodriguez Parra}
\affiliation{Universidad Michoacana de San Nicolás de Hidalgo, Morelia, Mexico}

\author{D.~Rosa-González}
\affiliation{Instituto Nacional de Astrof\'{i}sica, Óptica y Electrónica, Puebla, Mexico}

\author{M.~Roth}
\affiliation{Los Alamos National Laboratory, Los Alamos, NM, USA}

\author{H.~Salazar}
\affiliation{Facultad de Ciencias F\'{i}sico Matemáticas, Benemérita Universidad Autónoma de Puebla, Puebla, Mexico}

\author{A.~Sandoval}
\affiliation{Instituto de F\'{i}sica, Universidad Nacional Autónoma de México, Ciudad de Mexico, Mexico}

\author{J.~Serna-Franco}
\affiliation{Instituto de F\'{i}sica, Universidad Nacional Autónoma de México, Ciudad de Mexico, Mexico}

\author{M.~Shin}
\affiliation{Department of Physics, Sungkyunkwan University, Suwon 16419, South Korea}

\author{A.J.~Smith}
\affiliation{Department of Physics, University of Maryland, College Park, MD, USA}

\author{Y.~Son}
\affiliation{University of Seoul, Seoul, Rep. of Korea}

\author{R.W.~Springer}
\affiliation{Department of Physics and Astronomy, University of Utah, Salt Lake City, UT, USA}

\author{O.~Tibolla}
\affiliation{Universidad Politecnica de Pachuca, Pachuca, Hgo, Mexico}

\author{I.~Torres}
\affiliation{Instituto Nacional de Astrof\'{i}sica, Óptica y Electrónica, Puebla, Mexico}

\author{L.~Villaseñor}
\affiliation{Facultad de Ciencias F\'{i}sico Matemáticas, Benemérita Universidad Autónoma de Puebla, Puebla, Mexico}

\author{X.~Wang}
\affiliation{Department of Physics, Missouri University of Science and Technology, Rolla, MO, US}

\author{Z.~Wang}
\affiliation{Department of Physics, Missouri University of Science and Technology, Rolla, MO, US}

\author{I.J.~Watson}
\affiliation{University of Seoul, Seoul, Rep. of Korea}

\author{H.~Wu}
\affiliation{Dept. of Physics and Wisconsin IceCube Particle Astrophysics Center, University of Wisconsin{\textemdash}Madison, Madison, WI, USA}

\author{S.~Yu}
\affiliation{Department of Physics, Pennsylvania State University, University Park, PA, USA}

\author{X.~Zhang}
\affiliation{Institute of Nuclear Physics Polish Academy of Sciences, PL-31342 IFJ-PAN, Krakow, Poland}

\author{H.~Zhou}
\affiliation{Tsung-Dao Lee Institute \& School of Physics and Astronomy, Shanghai Jiao Tong University, 800 Dongchuan Rd, Shanghai, SH 200240, China}

\author{C.~de León}
\affiliation{Universidad Michoacana de San Nicolás de Hidalgo, Morelia, Mexico}

\nocollaboration{4}
\author[0000-0002-5152-2971]{Y. Li}
\affiliation{
GRAPPA Institute, University of Amsterdam, 1098 XH Amsterdam, The Netherlands} 

\author[0000-0003-4679-1058]{J. Gelfand}
\affiliation{New York University Abu Dhabi, P.O. Box 129188, Abu Dhabi, United Arab Emirates}

\author{K. Ross}
\affiliation{
The Department of Aerospace, Physics And Space Sciences, Florida Institute of Technology,
Melbourne, FL, 32901, USA}

\author{M. Keith}
\affiliation{
The Department of Physics and Astronomy, University of Manchester,
Manchester, M13 9PL, UK}

\correspondingauthor{Jordan Eagle}
\email{jordanlynneagle@gmail.com}
\correspondingauthor{Yu Chen}
\email{ychen@astro.ucla.edu}
\correspondingauthor{Ruo-Yu Shang}
\email{r.y.shang@gmail.com}
\correspondingauthor{Ramiro Torres-Escobedo}
\email{torresramiro350@sjtu.edu.cn}
\correspondingauthor{Youyou Li}
\email{y.li4@uva.nl}



\begin{abstract}

\hrev{We present a} new study on the MeV--TeV $\gamma$-ray origin of HESS~J1857+026 using data collected from the Fermi--LAT, VERITAS, and HAWC observatories. %in addition to recently reported LHAASO data 
%in a multiwavelength study. 
A spatial and spectral study of HESS~J1857+026 including radiative modeling of the MeV--TeV spectrum determines the likely dominant $\gamma$-ray origin as a pulsar wind nebula (PWN) powered by the energetic pulsar PSR~J1856+0245. The MeV--TeV spectrum is further characterized through basic evolutionary radiative modeling assuming a PWN origin to constrain the physical properties of the system such as the magnetic field strength and PWN age. The results of the PWN evolutionary model are consistent with the observational constraints of the system, finding an age of the system \hrev{between $\tau = [16,21]\,$kyr} and a magnetic field strength \hrev{between $B = [0.4,1.6]\,\mu$G}. These estimates support an evolved PWN scenario where the observed $\gamma$-ray emission is generated by the relativistic electrons \frev{inverse Compton scattering (ICS)} off local photon fields, \hrev{however the} low-energy ($E < 10\,$GeV) spectral component \hrev{could be dominated by} hadronic \hrev{emission originating from a} \rev{supernova remnant (SNR)}. For a PWN component above 10\,GeV, we measure the conditions for particle diffusion, finding that the local diffusion ($D(\text{50\,TeV}) \sim 10^{28}\,$cm$^{-2}$s$^{-1}$) is suppressed compared to the interstellar medium (ISM) value, in agreement with similar TeV PWNe. \hrev{By measuring the radial surface brightness profiles of the $\gamma$-ray source across multiple instruments, we} \rev{demonstrate} that the combined MeV--TeV spatial information is a powerful tool to \rev{constrain particle diffusion properties}.
\end{abstract}



\section{Introduction}
The majority of pulsar wind nebulae (PWNe) have been discovered in the radio or X-ray bands and an increasing number of discoveries are occurring in TeV $\gamma$-rays. The majority of the Galactic TeV source population is found to be PWNe as observed by imaging atmospheric Cherenkov telescopes \citep[IACTs, e.g.][]{tevcat2008,acero2013}. These nebulae are strongly magnetized, relativistic particle winds powered by a rapidly rotating neutron star. Synchrotron emission from relativistic electrons is observed from the majority of PWNe, from radio wavelengths to hard X-rays. Additionally, the same relativistic electrons are expected to scatter off local photon fields, resulting in inverse Compton \rev{scattering} (ICS) emission at $\gamma$-ray energies \citep{gaensler2006}. In recent years, the High Altitude Water Cherenkov (HAWC) \hrev{observatory} and the Large High Altitude Air Shower Observatory (LHAASO) have reported the detection of energetic Galactic sources that are capable of emitting photons above 10\,TeV, most of them associated to known PWNe \citep{albert2020,cao2021ultrahigh,cao2023first}.

In order to accurately determine the underlying particle spectrum, we must understand the influences introduced from the evolution of the PWN inside its host supernova remnant \citep[SNR,][]{reynolds_1984,gelfand2009dynamical}. At early stages, the PWN expands freely into the interior of the SNR until the reverse shock returns and crushes the PWN. An asymmetric morphology is often observed after the PWN has been crushed due to the inhomogeneous density the SNR expands into \citep[e.g.,][]{slane2018,eagle2022}. The crushed PWN will undergo compression, which may enable the central pulsar to exit the nebula. As a result, a relic nebula is left behind while the continuous injection of high-energy particles by the pulsar forms a fresh nebula concentrated close to the pulsar. 

\rev{In later evolutionary stages of the PWN, most electrons have cooled to energies at which the cooling time is as long as the time since their injection. These electrons will primarily radiate in the radio and GeV bands. The most recently injected particles have not significantly cooled and emit X-ray synchrotron and TeV IC photons. They are located close to the pulsar, on account of the short time available for propagation. } The resulting energy-dependent morphology is observed in several PWNe including HESS~J1825--137 \citep{principe2020}, HESS~J1303--631 \citep{j13032012}, G327.1--1.1 \citep{temim2015,eagle2022}, and MGRO~J1908+06 \citep{shang2024}. \hrev{See \cite{gaensler2006,sudoh2019tev,giacinti2020halo} for additional discussions on the evolution of PWNe.}  %On a spectral energy distribution (SED), the high-energy population manifests as the highest-energy emission found above the cooling break energy while the oldest particles (i.e., low-energy population) are present below this energy.  

%Finally, we have performed a similar investigation to the one we report here in \citet{shang2024}. 
%Using data from the Fermi--Large Area Space Telescope (LAT), VERITAS, and HAWC observatories, we characterize the GeV--TeV emission from MGRO~J1908+06, which was found to most likely be an evolved PWN powered by PSR~J1907+0602. 
In this paper, we report the investigation for the TeV PWN candidate HESS~J1857+026 using data from the Fermi--Large Area Space Telescope (LAT) and \hrev{TeV data from the} \hrev{Major Atmospheric Gamma Imaging Cherenkov (MAGIC) observatory} \citep{magic2014}, \hrev{the High Energy Stereoscopic System}, \citep[HESS,][]{hess_reichardt2015}, \rev{the Very Energetic Radiation Imaging Telescope Array System (VERITAS)}, and \rev{HAWC}. In the following section\rev{,} we provide an overview for the source of interest. In Section~\ref{sec:fermi_analysis}\rev{,} we present the data analysis of 15\,years of LAT data in the 300\,MeV--2\,TeV band. In Section~\ref{sec:veritas_analysis}\rev{,} we present the VERITAS data analysis in the \rev{0.3\,--\,10}\,TeV band and in Section~\ref{sec:hawc_analysis} we present the HAWC data analysis \hrev{for estimated energies  1\,--\,316\, TeV}. In Section~\ref{sec:mw}\rev{,} we perform broadband modeling and in Section~\ref{sec:pwn_physics} we describe the diffusion environment for the system based on observational constraints and the best-fit model results from Section~\ref{sec:mw}. In Section~\ref{sec:conclusion}\rev{,} we provide our conclusions.

\begin{figure}
\hspace{-0.65cm}
\includegraphics[width=1.07\linewidth]{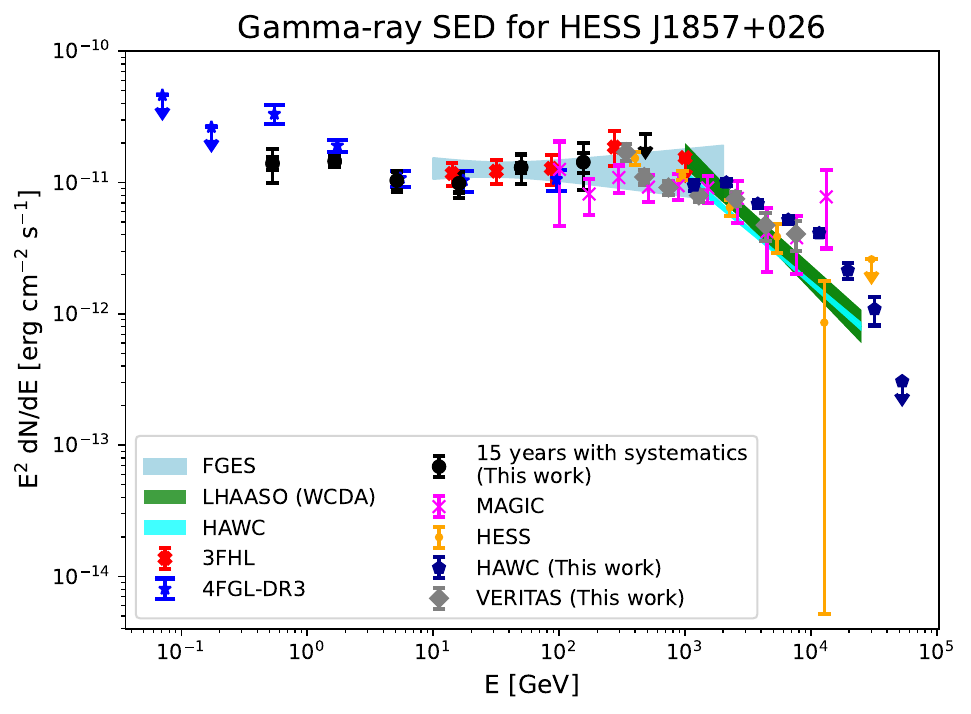}
\caption{The 50\,MeV to 30\,TeV $\gamma$-ray SED for HESS~J1857+026. The blue flux \hrev{stars} are from the Fermi--LAT 4FGL--DR3 catalog \citep{4fgldr3} and the light blue uncertainty flux band is from the Fermi--LAT FGES catalog \citep{ackermann2017}. The red-\hrev{filled X-points} are from the Fermi--LAT 3FHL catalog \citep{3fhl2017}. The green uncertainty flux band is from the 1--25\,TeV LHAAASO catalog data \citep{cao2023first} and the 1.3--32\,TeV cyan uncertainty flux band is from the 3HAWC survey \citep{albert2020}. The black \hrev{circles} are from the Fermi--LAT between 300\,MeV and 2\,TeV (this work, see Section~\ref{sec:fermi_analysis}), the grey \hrev{diamonds} are from VERITAS (this work, see Section~\ref{sec:veritas_analysis}), the dark blue \hrev{hexagons} are new HAWC data (this work, see Section~\ref{sec:hawc_analysis}), the orange \hrev{points} are from HESS \citep{hessgps2018} and the pink \hrev{X-points} from MAGIC \citep{magic2014}. The Fermi--LAT systematics are from \citet{eagle_2022}.}
\label{fig:gammaray_sed}
\end{figure}

\section{HESS~J1857+026}
%\lipsum[1-2]
HESS~J1857+026 was first detected as an unidentified, extended, very high energy (VHE, $E > 100\,$GeV) $\gamma$-ray source by HESS \citep{aha2008}. It \rev{became} a PWN candidate after the radio discovery of a pulsar, PSR~J1856+0245, that is located at the TeV emission peak and has a spin-down luminosity ($\dot{E} \sim 5 \times 10^{36}\,$erg s$^{-1}$) that can explain the observed TeV luminosity \citep{hessels2008}. The pulsar is reported as ``Vela-like'' with a characteristic age $\tau_c = 10-30\,$kyr, a spin period of 81\,ms, and a large dispersion measure (DM) of 622 cm$^{-3}$ pc which corresponds to a derived distance of $\sim 9\,$kpc assuming the NE2001 electron density model \citep{hessels2008}. \citet{hessels2008} reported the faint detection of an ASCA X-ray source coincident to the pulsar, detected only in the hard X-ray range (2--10\,keV), which may indicate the presence of a compact X-ray PWN counterpart. The unabsorbed flux of the potential X-ray PWN is measured as $1.6 \times 10^{-13}$\,erg cm$^{-2}$ s$^{-1}$ in 2--10\,keV \citep{hessels2008}. Chandra and XMM-Newton clearly detect PSR~J1856+0245 in a 39\,ks ACIS-I observation and a 30\,ks EPIC pn observation, but no diffuse emission is visible, constraining the upper limit on the unabsorbed flux down to $5 \times 10^{-14}$\,erg cm$^{-2}$ s$^{-1}$ in 1--10\,keV \citep{rouss2012} for the immediate region around the pulsar. \rev{Assuming the source size in X rays is commensurate with that seen in the TeV band with HESS,} the upper limit becomes $2 \times 10^{-12}$\,erg cm$^{-2}$ s$^{-1}$ in 1--10\,keV \citep{rouss2012}. While the detection of the X-ray counterpart to the PWN is not confirmed, it is consistent with the observed X-ray morphology for many evolved or ``relic'' PWNe, which can be very faint or even absent in \frev{X-rays} due to the synchrotron cooling time of the highest energy electrons \citep{karg2013}. No thermal X-rays from the pulsar nor the SNR shell \hrev{have} been identified either, \rev{which} may be due to the high absorption in the direction of the pulsar \citep{nice2013}.

Point-like $\gamma$-ray emission coincident to HESS~J1857+026 is detected by the Fermi--LAT for $E>10\,$GeV \citep{rouss2012}, \rev{located 0.17\,$^\circ$ North of the pulsar}. A pulsation search around the radio pulsar position utilizing a radio ephemeris constructed from a series of observations finds no significant pulsation in the Fermi--LAT energy band. \citet{rouss2012} derive an upper limit on the \hrev{0.1--1\,GeV} flux at the pulsar location to be $\sim 3 \times 10^{-8}\,$ph cm$^{-2}$ s$^{-1}$, making it unlikely to be generating the observed GeV emission. In \rev{the 100\,MeV--100\,GeV band}, HESS~J1857+026 is well described by a simple power-law with a spectral index $\Gamma_\gamma \sim 1.5$. %at above the $5\,\sigma$ significance level. 
Assuming the distance to the PSR~J1856+0245 of 9\,kpc, this corresponds to a $\gamma$-ray luminosity of $L_{\text{100\,MeV--100\,GeV}}$ $\sim 2.5 \times 10^{35} \big(\frac{d}{9\text{kpc}}\big)^2\,$erg s$^{-1}$, or a $\gamma$-ray efficiency $\eta \times 100\% $ of $\sim 5\%$, a value typical of PWNe. A similar value is found for the TeV luminosity \citep[$\sim 3\%,$][]{hessels2008}. Extended $\gamma$-ray emission \rev{was} later detected by the Fermi--LAT for $E>10\,$GeV towards HESS~J1857+026 \citep{ackermann2017} and is currently reported as 4FGL~J1857.7+0246e in the latest Fermi--LAT comprehensive source catalog, the 4FGL-Data Release 3 \citep[DR3,][]{4fgldr3}. The source extent is modeled using a uniform disk with radius $r = 0.61\,^\circ$, much larger than the HESS extension $\sim 0.1\,^\circ$.

HESS~J1857+026 is also detected by the MAGIC IACT \citep{magic2014} $> 150\,$GeV %at a $12\,\sigma$ significance level and 
with a best-fit power-law spectral index $\sim 2.2$. The source, MAGIC~J1857.2+0263, is extended and fitted as an elliptical Gaussian with semi-major and -minor axes of $\sim 0.17\,^\circ$ and $\sim 0.06\,^\circ$\rev{, respectively}. The VHE emission in this region becomes visibly complex above 1\,TeV, where \frev{an additional} source is detected to the North of the region, MAGIC~J1857.6+0297. \cite{magic2014} \rev{performed} a molecular \hrev{gas} distribution study investigating the possibility of two sources in the MAGIC data and \rev{found} that both a hadronic (MAGIC~J1857.6+0297) and a leptonic (MAGIC~J1857.2+0263) contribution are plausible. Similar conclusions are also described in \cite{hess_reichardt2015}. %The large extension for the overlapping 4FGL~J1857.7+0246e supports this scenario. Did I write this??
A comprehensive $\gamma$-ray spectral energy distribution (SED) of HESS~J1857+026 displaying all available data is displayed in Figure~\ref{fig:gammaray_sed}. 

\begin{figure}
%\includegraphics[width=1.0\linewidth]{HI_petriella2021_placeholder.png}
%\caption{Nice HI image similar to Petriella 2021 showing the HI cavity along with the PSR location and flux contours for Fermi, VERITAS, and HESS.}
\includegraphics[width=0.9\linewidth]{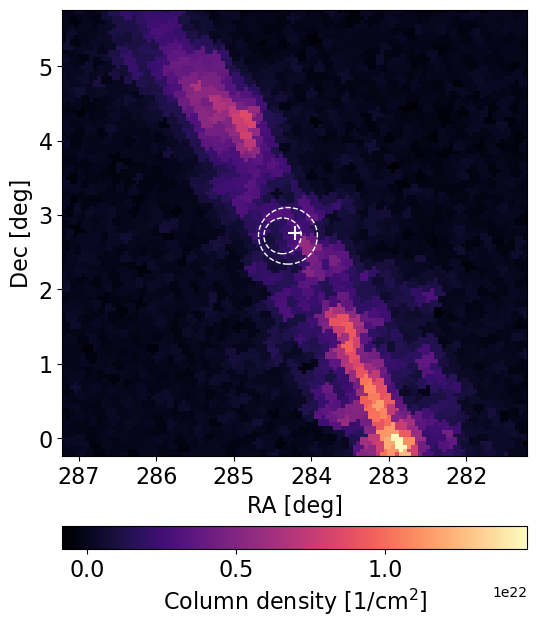}
\includegraphics[width=0.9\linewidth]{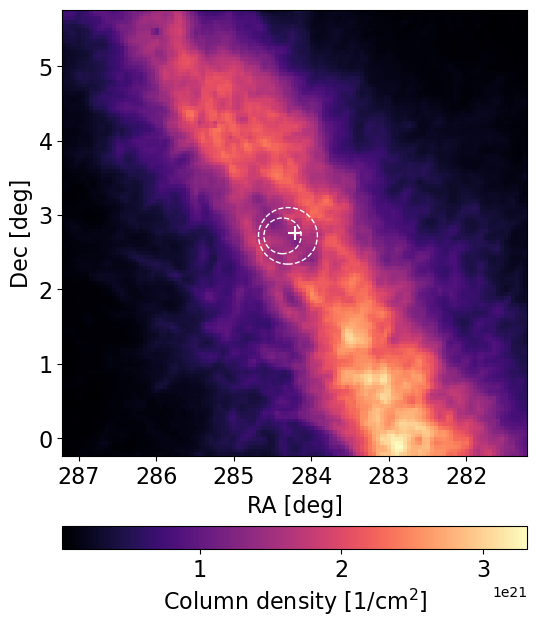}
\caption{
The particle column densities estimated from CO (top) and HI (bottom) emission in the region around HESS~J1857+026 in the velocity range 81\,km s$^{-1}$ \hrev{to} 102\,km s$^{-1}$, corresponding to a distance between 5.3\,kpc and 6.1\,kpc. 
The larger circle shows the $\gamma$-ray size from Fermi-LAT data (see Section~\ref{sec:fresults}). The smaller circle shows the $\gamma$-ray size from the VERITAS data (see Section~\ref{sec:veritas_analysis}). 
The white cross sign represents the location of PSR J1856+0245. 
The $^{12}$CO ($J=1-0$) data \citep{dame2001milky} are retrieved from the 1.2-m CO Survey Dataverse of the Smithsonian Astrophysical Observatory. The HI data are obtained from the Galactic Archive of the Arecibo L-band Feed Array %using the Arecibo 305-meter radio telescope 
\citep[GALFA,][]{peek2017galfa}.
}
\label{fig:cloud_map}
\end{figure}

The possible source confusion in the region was further explored by \citet{petriella2021} utilizing resolved radio observations for both neutral hydrogen (HI) and carbon monoxide (CO). The authors report, down to the noise level, the nondetection of a PWN around PSR~J1856+0245 in both 1.5\,GHz and 6.0\,GHz. Any hint of a radio SNR shell is not apparent either. The distance to the pulsar is re-derived based on the discovery of an HI cavity-like structure surrounding the bulk of TeV emission and pulsar, finding near and far distances of 5.5 and 8.3\,kpc respectively \citep{petriella2021}. The near distance of 5.5\,kpc is most compatible with the DM distance measured for PSR~J1856+0245 of 6.3\,kpc assuming the YMW16 electron density model, so 5.5\,kpc to PSR~J1856+0245 is preferred in \citet{petriella2021}. If the pulsar is indeed associated with the HI cavity, \rev{this is consistent with a} scenario where the cavity was generated by progenitor stellar winds and later by the expansion of the SNR. \citet{petriella2021} argue in favor of a single $\gamma$-ray source based on the spatial agreement between the HI cavity, a possible wind-blown superbubble, and the TeV emission from HESS~J1857+026. The lack of spatial coincidence between the TeV and CO emission peaks challenges a hadronic contribution for the TeV emission. Figure~\ref{fig:cloud_map} shows CO (top panel) and HI (bottom panel) emission in the velocity range 81\,km s$^{-1}$ \hrev{to} 102\,km s$^{-1}$, corresponding to a distance between 5.3\,kpc and 6.1\,kpc\footnote{\url{https://dataverse.harvard.edu/dataverse/rtdc}}. The GeV and TeV $\gamma$-ray emission are located in a region with low molecular gas density.
%A dominant leptonic origin therefore seems to be more likely, though a hadronic contribution remains possible. 

\hrev{Finally, a possible neutrino excess detected by IceCube has been reported for the HAWC counterpart at the $\sim$2\,$\sigma$ significance level \citep[2HWC~J1857+027,][]{neutrino2019,neutrino2021}. Further studies of the possible neutrino excess are needed to better determine the likelihood for the HAWC emission to be associated, and which the future KM3NeT detector is expected to enable.}

\section{Fermi--LAT data analysis}\label{sec:fermi_analysis}
The principal scientific instrument on the Fermi Gamma-ray Space Telescope is the LAT \citep{atwood2009}. The LAT instrument is sensitive to $\gamma$-rays with energies from 50\,MeV to $> 300$\,GeV \citep{4fgl2020}. The \rev{LAT has an} instantaneous field of view \rev{of} $\sim 2.4$ steradian and has been continuously surveying the entire sky every 3\,hours since beginning operation in 2008 August. 

We analyze 15\, years (from 2008 August to 2023 July) of Pass~8 \texttt{SOURCE} class data \citep{atwood2013,pass82018} between 300\,MeV and 2\,TeV. Photons detected at zenith angles larger than 100\,$^\circ$ were excluded to limit the contamination from $\gamma$-rays generated by cosmic ray (CR) interactions in the upper layers of Earth's atmosphere. We perform a binned likelihood analysis of the best quality of reconstructed photon events (\texttt{PSF3} type) using the latest Fermitools package\footnote{\url{https://fermi.gsfc.nasa.gov/ssc/data/analysis/software/}} (v.2.2.11) and FermiPy Python~3 package \citep[v.1.2.0][]{fermipy2017}. We fit the square 10$\,^\circ$ region of interest (ROI) in equatorial coordinates using a pixel bin size $0.05\,^\circ$ and 8 bins per decade in energy (31 total energy bins). The $\gamma$-ray sky for the ROI is modeled from the latest comprehensive Fermi--LAT source catalog based on 12\,years of data, 4FGL--DR3 \citep{4fgldr3} for point and extended sources\footnote{{\url{https://fermi.gsfc.nasa.gov/ssc/data/access/lat/12yr_catalog/}.}} that are within 15\,$^\circ$ of the ROI center, as well as the latest Galactic diffuse and isotropic diffuse templates (\texttt{gll\_iem\_v07.fits} and \texttt{iso\_P8R3\_SOURCE\_V3\_PSF3\_v1.txt}, respectively)\footnote{LAT background models and appropriate instrument response functions: \url{https://fermi.gsfc.nasa.gov/ssc/data/access/lat/BackgroundModels.html}.}.

With the source model described above, we allow the background components and sources within distances from the ROI center $\leq3.0$\,$^\circ$ to vary in normalization. The test statistic (TS) value quantifies the significance for a source detection with a given set of location and spectral parameters and the significance of such a detection can be estimated by taking the square root of the TS value for 1 degree of freedom \citep[DOF,][]{mattox1996}. The TS value is defined to be the natural logarithm of the ratio of the likelihood of one hypothesis (e.g., presence of one additional source) and the likelihood for the null hypothesis (e.g., absence of source):
\begin{equation}
  TS = 2 \times \ln\left({\frac{{\mathcal{L}_{1}}}{{\mathcal{L}_{0}}}}\right)
    \label{equation:ts}
\end{equation}
TS values $>25$ correspond to a detection significance $> 4 \sigma$ for 4 DOF.

\begin{figure*}
\gridline{
\fig{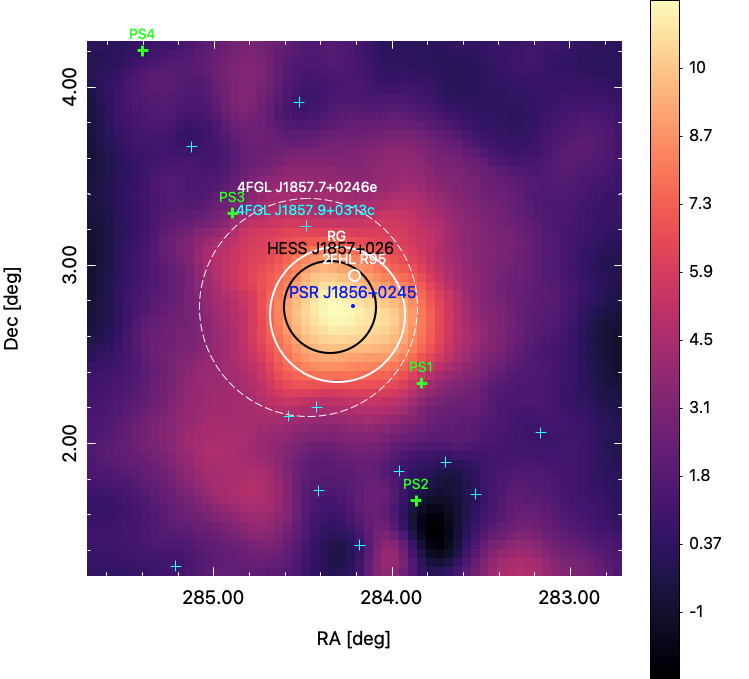}{0.51\textwidth}{(a)}
\fig{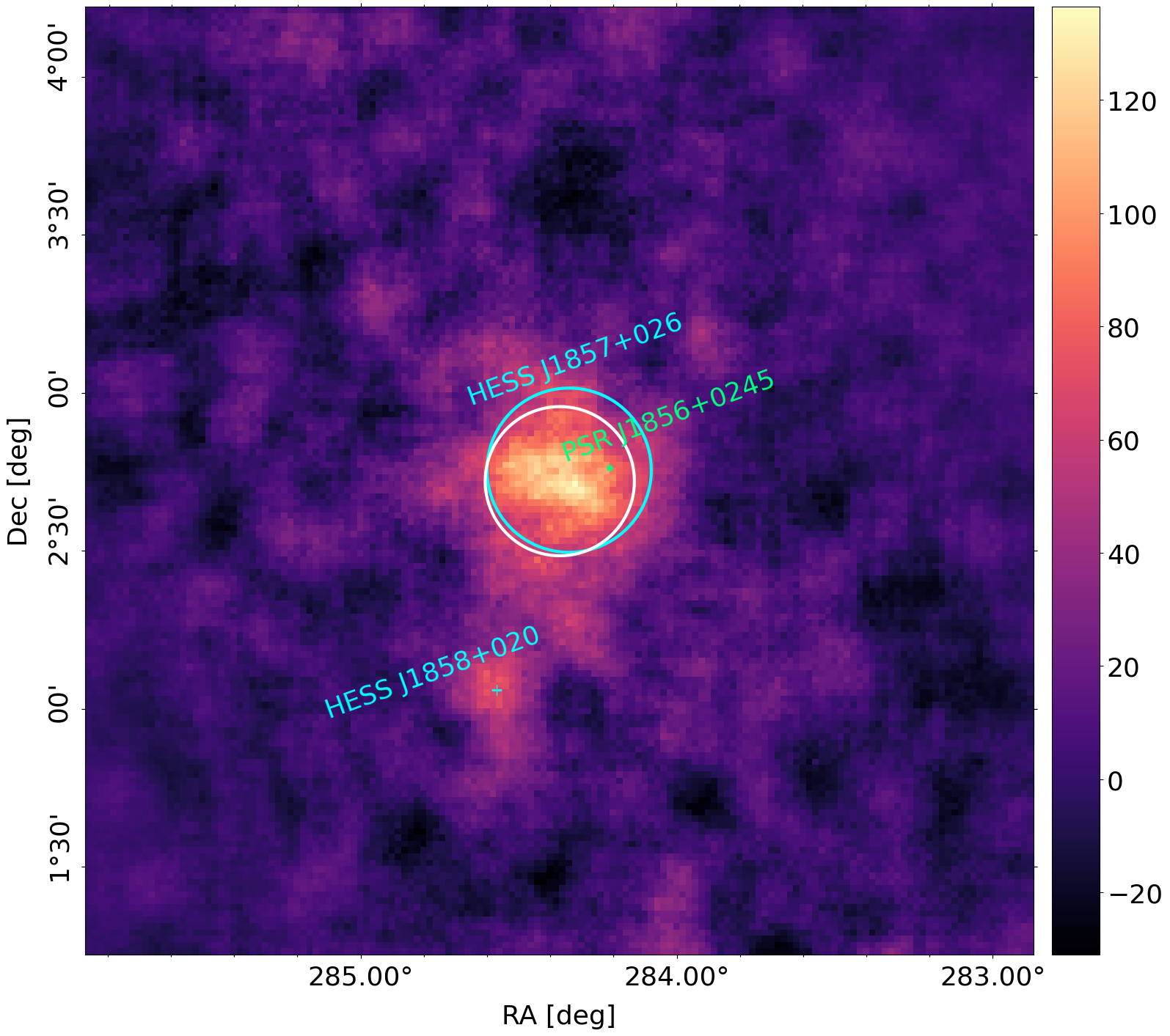}{0.48\textwidth}{(b)}
% \fig{hist_real_diff_skymap_total_excess_counts_map3x3_magma}{0.5\textwidth}{(b)}
}
\caption{
{\it Left:} A 3\,$^\circ \times 3\,^\circ$ excess counts map of Fermi--LAT data with energy between 300\,MeV and 2\,TeV. Unrelated 4FGL sources are in cyan. Additional point sources are labeled in \apjrev{green}. 4FGL~J1857.7+0246e (white dashed circle) is replaced in the model with the radial Gaussian source (RG) marked as the solid white circle. The 95\% positional uncertainty for 2FHL~J1856.8+0256 is shown as the smaller white circle, see text for details. {\it Right:} A 3\,$^\circ \times 3\,^\circ$ excess counts map with VERITAS data in the energy range 0.3--10\,TeV smoothed with a correlation radius of 0.1\,$^\circ$. The Gaussian extension of the VERITAS emission associated with HESS~J1857+026 in this work is shown as the solid white circle. Unrelated VERITAS emission is seen to the south, corresponding to HESS~J1858+020, and is marked with a cyan cross. {\it Both Panels:} The HESS~J1857+026 extension in the HGPS catalog \citep{hessgps2018} is indicated as a black circle on the left and a cyan circle on the right. The X-ray position of PSR~J1856+0245 is marked with a smaller circle that has the approximate size of the compact PWN (blue color on the left and green color on the right).
}
\label{fig:fermi_veritas_excess_maps}
\end{figure*}

Figure~\ref{fig:fermi_veritas_excess_maps} (a) displays the 300\,MeV -- 2\,TeV excess counts map centered on PSR~J1856+0245 and is generated from the best-fit global source model which replaces 4FGL~J1857.7+0246e with a radial Gaussian slightly positionally offset and with a smaller extension than the 4FGL source (see details below in Section~\ref{sec:fresults}). The excess is significantly extended and corresponds to a 10\,$\sigma$ detection significance. 

\subsection{Fermi--LAT Data Analysis Results}\label{sec:fresults}
To model the $\gamma$-ray emission coincident to HESS~J1857+026, we first model significant residual excesses in the ROI that are unrelated emission but may be sources of contamination if not included. There are 4 such sources which we model as point sources assuming power-law spectra and are labeled as PS~1--4 in Figure~\ref{fig:fermi_veritas_excess_maps} (a). We then remove 4FGL~J1857.7+0246e, the 4FGL counterpart to HESS~J1857+026, to re-characterize all emission in the immediate region. We test a point source at the pulsar location R.A., Dec. = 284.212\,$^\circ$, +2.763\,$^\circ$ (J2000). 
We set the spectrum to a power law characterized by
\begin{equation}
  \frac{dN}{dE} = N_{0} \big(\frac{E}{E_0}\big)^{-\Gamma}
  \label{equation:pl}
\end{equation}
where $E_0$ is set to 1000\,MeV. We allow the spectral index and normalization to vary. We localize the point source with \texttt{GTAnalysis.localize} to find the best-fit position and uncertainty. %The localized position for the new $\gamma$-ray source is offset by 0.04\,$^\circ$ from the exact position of PSR~J1856+0245, corresponding to the nearby TS peak and has R.A., Dec. = 284.23\,$^\circ$, +2.70\,$^\circ$ (J2000). The corresponding 95\% positional uncertainty radius is $r=0.07 ^\circ$. The TS of the $\gamma$-ray source is 135 at this location with a spectral index $\Gamma_\gamma = 2.60 \pm 0.09$.
\begingroup
\begin{table*}[!htb]
\centering
\begin{tabular}{cccccc}
\hline
\hline
\ Spatial Template & TS & TS$_{\text{ext}}$ & (R.A., Dec.) ($^\circ$, J2000) & $r_{68}$ ($^\circ$) & 95\% U.L. ($^\circ$) \
\\
\hline
Point Source & 134.8 & -- & 284.23, +2.70 & -- & -- \\
Radial Disk & 285.9 & 121.8 & 284.28, +2.72 & 0.30 $\pm$ 0.02 & 0.34 \\
Radial Gaussian & 327.8 & 128.5 & 284.30, +2.72 & 0.38 $\pm$ 0.04 & 0.45 \\
\hline
\hline
\end{tabular}
\caption{Summary of the best-fit parameters and the associated statistics for each spatial template used in our Fermi--LAT analysis. \hrev{Uncertainties on the extension are the 1\,$\sigma$ statistical errors.} \frev{The radius is quoted for 68\% containment and corresponds to $\sigma = \frac{r}{1.51} = 0.25\,^\circ$ for the Gaussian model.} The final column represents the 95\% upper limit for the extension.}
\label{tab:extent}
\end{table*}
\endgroup
We perform extension tests on the best-fit point source utilizing \texttt{GTAnalysis.extension} and the two spatial templates supported in the FermiPy framework, the radial disk and radial Gaussian templates. Both of these extended templates assume a symmetric 2D shape with width parameters radius and sigma, respectively. We allow the position and spectral parameters to vary when finding the best-fit spatial extension. The best-fit parameters for the extension tests are presented in Table~\ref{tab:extent}. 

\begin{figure}
\centering 
\includegraphics[width=1.0\linewidth]{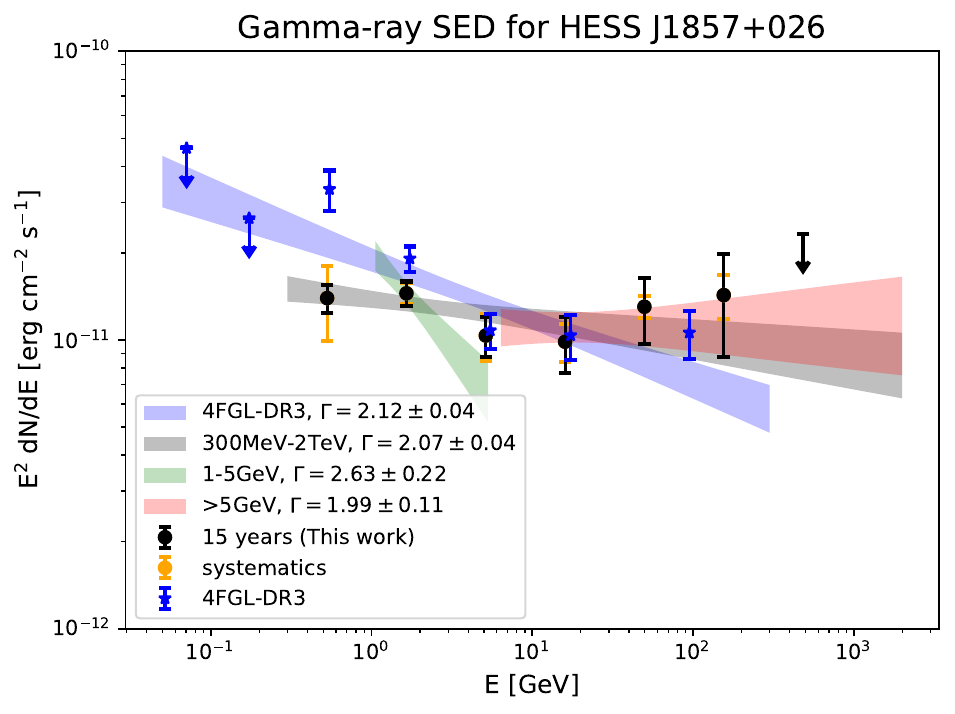}
\caption{The Fermi--LAT SED for the extended source reported in Section~\ref{sec:fermi_analysis} in three energy bands: 1--5\,GeV (green), 5\,GeV--2\,TeV (red), and 300\,MeV--2\,TeV (black) and compared to the 4FGL--DR3 (blue). }
\label{fig:dr3_this_work}
\end{figure}

The best-fit spatial template is determined by maximizing \rev{TS$_{\text{ext}} = 2 \times \ln\big(\frac{\mathcal{L}_{\text{ext}}}{\mathcal{L}_{\text{ps}}}\big)$}. The radial Gaussian template is found to provide the best-fit with TS$_{\text{ext}}$ = 128 and an extension \hrev{$r = 0.38^{\circ} \pm 0.04^{\circ}_{\rm stat}$} at the location R.A., Dec. = 284.30\,$^\circ$, +2.72\,$^\circ$ (J2000) with a 95\% uncertainty on the Gaussian centroid that is 0.07\,$^\circ$. The TS and best-fit power-law index for the radial Gaussian source are 327.8 and $\Gamma_\gamma = 2.07 \pm 0.04$, respectively. The integrated energy flux for the Gaussian source in the 300\,MeV--2\,TeV energy band is 9.6 $\pm$ 1.0 $\times 10^{-11}$\,erg cm$^{-2}$ s$^{-1}$. Systematic uncertainties introduced by the choice in the background model and from the LAT instrument performance are included in the flux measurements, adopting those measured in \citet{eagle_2022}. The systematic uncertainties dominate over the statistical errors for the lowest energy bin ($E<1$\,GeV), see Figure~\ref{fig:dr3_this_work}. 

Finally, we perform a pulsation search using \texttt{pint-pulsar} \citep{pintpulsar} in the 300\,MeV--2\,TeV energy range and an updated timing solution for PSR~J1856+0245 from the Jodrell Bank Observatory, covering \rev{the} 22 May 2006 to 16 Nov 2020 time span. No pulsations are detected in the LAT data. 

The results reported here are in agreement with previous Fermi--LAT analyses \citep[][see also Figure~\ref{fig:gammaray_sed}]{2fhl2016,ackermann2017,3fhl2017,eagle_2022,4fgldr3}. There is some evidence for an additional hard-spectrum source embedded within the $\gamma$-ray signal, first revealed in the Fermi--LAT 2FHL catalog with the detection of 2FHL~J1856.8+0256 which has a best-fit photon index $\Gamma_\gamma = 2.0 \pm 0.38$ above 50\,GeV \citep{2fhl2016}. %To explore this, we add another point source assuming a simple power-law spectrum at the 2FHL source position and localize it in addition to considering the best-fit Gaussian source. A marginal improvement to the fit is found, 
When added to the source model, the additional source is localized just North of the 2FHL position (R.A., Dec.) = (284.43\,$^\circ$, +2.98\,$^\circ$) and has $\text{TS} = 9$, a $< 3\,\sigma$ detection, and a photon index $\Gamma_\gamma = 1.8 \pm 0.26$. Given the marginal improvement in the fit, we cannot determine whether a second spectral component is present, but it also cannot be ruled out. %If a second harder spectral component is confirmed, this would imply the presence of a young, compact PWN concentrated near the central pulsar. 

\begin{figure}[b]
\centering 
\includegraphics[width=1.0\linewidth]{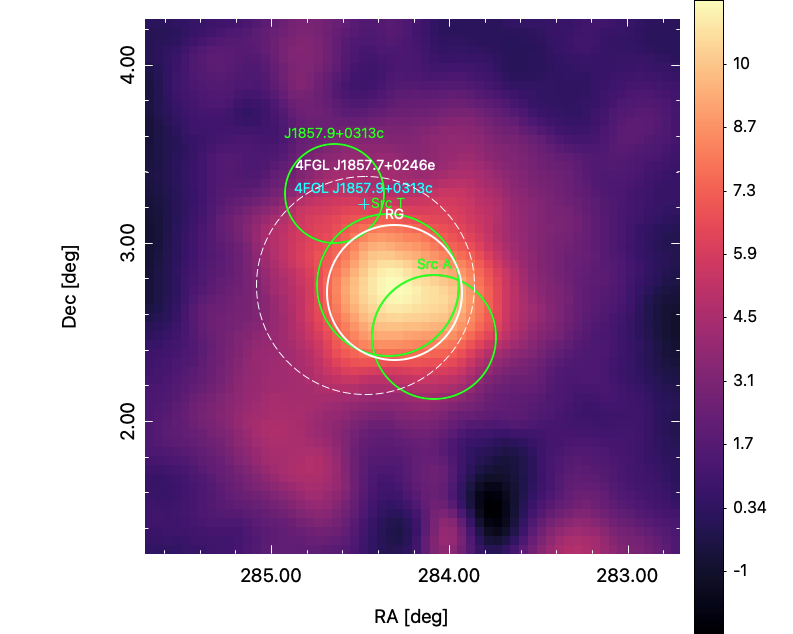}
\caption{\hrev{Same as Figure~\ref{fig:fermi_veritas_excess_maps}(a) but comparing the three source models discussed in the main text: the 4FGL, the radial Gaussian model we report in Table~\ref{tab:extent}, and the model comprising three extended sources presented by \citet{guo2023}. The 4FGL source is displayed as the white dashed circle, our best-fit radial Gaussian (``RG'') source as the solid white circle, and the three extended sources reported by \citet{guo2023} in green. The northern most extended source in \citet{guo2023} replaces the point source 4FGL~J1857.9+0313c, which is shown as the cyan cross. See text for details.}}
\label{fig:guo_compare_map}
\end{figure}

Recent work \frev{\citep{guo2023}} investigated the GeV and TeV morphological and spectral origins, finding that three GeV extended sources can explain the Fermi--LAT data: two (\frev{named} SrcA and 4FGL~J1857.9+0313c) are brighter in \rev{lower energies} ($E < 3\,$GeV) and the other (\frev{named} SrcT) is brighter in \rev{higher energies} ($E > 10\,$GeV). %SrcT is reported to have energy-dependent morphology above 10\,GeV with a decreasing size concentrating towards PSR~J1856+0245. %, in agreement with the results of the Fermi--LAT, VERITAS, and HAWC analyses here. 
%\frev{One of the three extended GeV sources, modeled as a 2D disk, replaces} 4FGL~J1857.9+0313c \frev{and} falls to the north of \frev{4FGL~J1857.7+0246e (see} Figure~\ref{fig:guo_compare_map}). \frev{With a} disk radius $r \sim0.29\,^{\circ}$, this source \frev{models part of the} northern emission of 4FGL~J1857.7+0246e. \frev{The second extended source, SrcA, has a} Gaussian size $\sim 0.35\,^{\circ}$ and falls to the south \frev{of 4FGL~J1857.7+0246e, such that SrcA partially models the southern region of 4FGL~J1857.7+0246e}. \frev{The third source, SrcT,} is a Gaussian source with $\sim 0.40\,^{\circ}$ size that would \frev{best} correspond to 4FGL~J1857.7+0246e, \frev{which matches the} 2D Gaussian \frev{we report here (Table~\ref{tab:extent})}. SrcT has a hard photon index $\Gamma_\gamma = 1.74 \pm 0.07$, while \frev{the two other extended sources}, SrcA and the extended source modeling 4FGL~J1857.9+0313c, have softer photon indices, $\Gamma_\gamma \sim 2.73$ and $\Gamma_\gamma \sim 2.55$, respectively. %The Gaussian morphology of SrcT agrees with the 2D Gaussian reported here and is plausibly the same high-energy source from \citet{guo2023}. 
%The spectral index of SrcT is harder than the \hrev{spectral index of the} 2D Gaussian we report, but agrees with the additional hard-spectrum source.
We test the \frev{presence of three extended sources} presented by \citet{guo2023} (\frev{i.e.,} Src~A, an extended version of 4FGL~J1857.9+0313c, \frev{and Src~T}, \apjrev{see Figure~\ref{fig:guo_compare_map}}). A single extended source, \frev{corresponding to the 2D Gaussian reported here (Table~\ref{tab:extent}),} remains to be the best-fit. \frev{The results would match closest to Src~T of the model presented in \citet{guo2023}.} This does not rule out multiple emission components to \frev{a single} extended source, \frev{so the power-law source spectrum is additionally measured in two energy bands chosen to be similar to \citet{guo2023}, 1--5\,GeV and 5\,GeV--2\,TeV, and compared to the spectral results found using energies between 300\,MeV and 2\,TeV above.} Compared to the 300\,MeV--2\,TeV spectral index value $\Gamma = 2.07 \pm 0.04$, a softer spectral index is measured between 1--5\,GeV for the extended source, $\Gamma = 2.63 \pm 0.22$, and a harder index for $>5$\,GeV, $\Gamma = 1.99 \pm 0.11$. We plot the different spectra in Figure~\ref{fig:dr3_this_work}. In conclusion, two spectral components may contribute to the extended Fermi--LAT emission: a soft component peaking \hrev{at} $E \lesssim 10$\,GeV and a hard component peaking \hrev{at} $ E \gtrsim 10\,$GeV. It is thus possible that a hadronic contribution is present, which would agree with the results of \citet{guo2023}. Namely, \rev{\citet{guo2023} argue that} SrcA and \frev{an} extended 4FGL~J1857.9+0313c are the likely hadronic contributors and SrcT the leptonic contributor, of which all three are encompassed in the single extended source found here. However, it is also possible that the spectral components \frev{visible in Figure~\ref{fig:dr3_this_work}} are the result of two different electron populations from the same, evolved PWN, which is commonly observed for several late-stage PWNe \citep[e.g.,][]{hinton2011,temim2015} and is consistent with the energy-dependent morphology of SrcT. 

%Finally, we note that \frev{two sources in the constructed global source model,} 4FGL~J1857.6+0212 and 4FGL~J1858.3+0209, correspond to HESS~J1858+020. These sources are likely unrelated to HESS~J1857+026 and are instead plausibly associated with SNR~G35.6--0.4 or an HII region \citep{zhang2022}. HESS~J1858+020 is \frev{brightly detected by the Fermi--LAT, VERITAS, and HAWC; } \frev{therefore it is} considered in the Fermi--LAT, VERITAS, and HAWC fits. 

\section{VERITAS data analysis}\label{sec:veritas_analysis}

VERITAS consists of four IACTs located at the Fred Lawrence Whipple Observatory \citep{weekes2002veritas}. 
Each telescope has a 12-m diameter reflector and a camera of 499 photomultiplier tube (PMT) pixels, covering a field of view of $3.5\,^{\circ}$. VERITAS is most sensitive to photons in the energy range 80\,GeV to 30\,TeV, with an optimal angular resolution of $0.08\,^{\circ}$ ($68\%$ containment radius) at 1\,TeV. The VERITAS data used in this study \rev{were} collected from 2008 to 2016.
After data quality selection, an effective total of 30 hours of exposure is available around the location of HESS~J1857+026.

%\begingroup
%\renewcommand*{\arraystretch}{1.0}
\begin{table*}[t!]
%\begin{flushleft} 
%\scalebox{1.0}{
\centering
\begin{tabular}{cccccc}
\hline
\hline
Source & Component & Parameter & Value & Uncertainty & Unit
\\
\hline
HESS~J1857+026  &Spatial & $\sigma$& 0.236& 0.017& deg ($^\circ$)\\
                &        & R.A.     & 284.369& 0.016& deg ($^\circ$)\\
                &        & Dec.     & 2.721& 0.020& deg ($^\circ$)\\
                &Spectral&$N_{0}$   & 5.59$\times 
 10^{-12}$& 4.5$\times 10^{-13}$& TeV$^{-1}$s$^{-1}$cm$^{-2}$ \\
                &        &$\Gamma$  & 2.48& 0.06& --\\
% \hline
% HESS J1858+020  &Spatial & $\sigma$& 0.13& 0.037& deg ($^\circ$)\\
%                 &        & R.A.     & 284.58                  & 0.023& deg ($^\circ$)\\
%                 &        & Dec.     & 2.06& 0.029& deg ($^\circ$)\\
%                 &Spectral&$N_{0}$   & 1.2$\times 10^{-12}$& 3.2$\times 10^{-13}$& TeV$^{-1}$s$^{-1}$cm$^{-2}$ \\
%                 &        &$\Gamma$  & 2.0                     & 0.11& --\\
\hline
\hline
\end{tabular}
\caption{Optimized parameters for HESS~J1857+026 in the VERITAS 0.3--40\,TeV source model. \hrev{$\sigma$ corresponds to the Gaussian width. The uncertainties listed here are statistical only.}}
\label{tab:VTS_gammapy_model_fit}
%\end{flushleft}
\end{table*}
%\endgroup

%HESS J1857+026 is an extended source in the TeV band.
The VERITAS event reconstruction and data reduction are achieved through a VERITAS analysis package Eventdisplay \citep{maier2017eventdisplay,Maier_Eventdisplay_An_Analysis_2024}. A boosted decision tree (BDT) score is assigned to each reconstructed event based on a machine-learning method that is trained to distinguish between CR and $\gamma$-ray showers \citep{KRAUSE2017BDT}. A cut based on the BDT score is applied to remove most of the CR background. The %background reduced 
remaining events, %are packed into DL3 files
combined with instrument response functions (IRFs), are converted to a standardized gamma-ray data format\footnote{\url{https://github.com/open-gamma-ray-astro/gamma-astro-data-formats}} using the pipeline described in \citet{V2DL3_-_VERITAS_2023}. %we use a tool\footnote{\url{https://github.com/VERITAS-Observatory/gammapy-tools}} developed by VERITAS members to 
We create background templates\footnote{\url{https://github.com/VERITAS-Observatory/gammapy-tools}} for each observing run that include CR rates in each spatial bin (offset from the center of the camera) and in each energy bin in the field of view (FOV). Each template is created with `off runs' that closely match the observing conditions (elevation, azimuth angle, night sky background rates, etc.) and have the $\gamma$-ray sources masked. In each spatial and energy bin, the counts of the ``off runs" are averaged to yield a mean rate value that is exposure corrected. The final background templates are then coupled to the appropriate IRFs. 
%The cosmic ray background estimation for the source is provided by the Low-rank Perturbation method (LPM), which is described in \textbackslash{}citet\{shang2024\}. The LPM utilizes the information of cosmic ray-like events that failed the \$\gamma\$-ray event selection from the observations to derive a background estimation for \$\gamma\$-ray-like background events. An independent background method is used as a cross-check for the LPM analysis. The results of the cross-check analysis is provided in the Appendix, Section\~\ref\{sec:gammapy\_ana\}.

We use the open-source software Gammapy \citep{gammapy:2023, gammapy_zenodo_v1p3} to perform the high-level data analysis. The VERITAS data are divided into logarithmically spaced energy bins (10 per decade) from 100\,GeV to 40\,TeV and fill a square ROI of 5\,$^\circ$ centered on the pulsar with a 0.02\,$^\circ$ spatial binning. We also note that the statistics are low for events with energies greater than 10 TeV. Therefore, we focus on the study of the source below 10\,TeV. Even though a global fit of the source model is still
performed including those higher energy bins, it was found to have negligible effect on the optimized results within uncertainties. The FOV background method \citep{berge_background_2007} is used that fits the background template in each run with a normalization and a tilt parameter to match the event counts simultaneously in spatial and energy bins after masking out the region of interest. The fitted background is then fixed in the following procedure. 

An excess counts map %\rev{\st{over the background}} 
in the energy range 0.3--10\,TeV is shown in Figure \ref{fig:fermi_veritas_excess_maps} (b). The $\gamma$-ray excess can be attributed to two sources, HESS~J1857+026 (to which we assign the additional name VER~J1857+027, corresponding to the VERITAS detection) and HESS~J1858+020, \frev{which comprise the global source model for the ROI as seen by VERITAS}. To characterize the morphological and spectral properties of each source, we use a 3D binned maximum-likelihood method assuming each source follows a simple power-law spectrum (Eqn.~\ref{equation:pl}, fixing $E_0 = 1\,$TeV) and has a 2D symmetric Gaussian spatial shape. %This is taken as the benchmark model. 
%The reference energy ${E_0}$ is fixed to 1\,TeV. 
The source locations are allowed to vary within $\pm$\,0.2\,$^\circ$ from the initial positions in each equatorial direction. The initial position of HESS~J1857+026 is chosen to be the pulsar location while the initial position of HESS~J1858+020 is chosen to be the \hrev{cataloged} location \citep{hessgps2018}. %, to ensure well-behaved convergent fitting results. 
%The initial test positions of HESS~J1857+026, and HESS~J1858+020 are R.A., Dec. = 284.212\,$^\circ$, +2.763\,$^\circ$(the pulsar location), and R.A., Dec. = 284.57\,$^\circ$, +2.06\,$^\circ$ (\citet{hessgps2018}), respectively. 
%A test statistic (TS) is defined in a similar way to the Fermi-LAT analysis presented in section \ref{sec:fermi_analysis} to evaluate the significance of the fit model of interest compared to an alternative hypothesis. $TS = 2 \times \ln\big({\frac{{\mathcal{L}_{1}}}{{\mathcal{L}_{0}}}}\big)$ where ${L}_{1}$ and ${L}_{0}$ are the likelihood of the model of interest, and the alternative hypothesis, respectively. 
The best-fit source model is chosen by maximizing the TS value following Eqn~\ref{equation:ts}. 

%The TS value of the benchmark model against a model that only includes HESS J1858+020 is 250.70, which 
The TS of the VERITAS emission modeled by HESS~J1857+026 corresponds to a 15.2 $\sigma$ detection.
% of HESS~J1857+026 with 5 additional degree of freedom. 
%We have also tested a model with an elongated Gaussian shape of HESS~J1857+026 against that with a symmetric Gaussian and a TS value of 3.57 is obtained, which is an insignificant improvement given 2 extra parameters.
%Therefore, the benchmark model is used to extract the spectrum. 
The optimized spatial and spectral parameters for the total source model of HESS~J1857+026 are displayed in Table~\ref{tab:VTS_gammapy_model_fit}. The Gaussian \frev{width} of HESS~J1857+026 is 0.236$\pm 0.017\,^\circ$ with a centroid that is 0.16\,$^\circ$ offset %sqrt((284.37-284.212)^2+(2.72-2.763)^2)
from the initial position (chosen to be the pulsar location). The power-law index for the spectrum of HESS~J1857+026 is $\Gamma = 2.48\pm0.06$. The 0.3--10\,TeV VERITAS spectrum is shown in Figure~\ref{fig:gammaray_sed}. The analysis results are cross-checked independently using the low-rank perturbation method for background estimation \citep[for details see][]{shang2024}.

\section{HAWC data analysis}\label{sec:hawc_analysis}
The  \hrev{HAWC} $\gamma$-ray observatory surveys the very high energy sky \hrev{for energies $>$1 TeV}. In the third HAWC Source Catalog \citep[3HWC,][]{albert2020}, the point source 3HWC~J1857+027 is reported with a detection significance $27.6\,\sigma$, a power-law spectral index \rev{$2.83\pm0.03_{-0.03}^{+0.10}$} in the 1.3--31.8\,TeV energy range, and is 0.14$\,^\circ$ away from HESS~J1857+026. We use the latest \textit{Pass 5} dataset, which comprises 2860 days from June 2015 to January 2024, an increase of \hrev{$\sim539$} \hrev{days} from previous work \citep{Albert_2024_crab}, \rev{to analyze emission associated with 3HWC~J1857+027}. This dataset also includes gamma/hadron cuts improved by two machine learning techniques, a multilayer perceptron and a convolutional neural network \citep{alfaroHAWCPerformanceEnhanced2025}. The data is divided into bins according to the fraction of PMTs that are triggered in each shower event, which are further subdivided into 12 quarter-decade \hrev{bins of estimated} energy covering the 0.316--316\,TeV range. This is performed using the neural network (NN) method presented in \cite{hawc2019}, which uses an artificial NN with a multilayer perceptron structure, with two hidden structures and a logistic activation function, to estimate the $\gamma$-ray photon primary energy based on parameters that are part of the standard HAWC event reconstruction.

\begin{figure*}[t]
    \centering
        \includegraphics[width=0.33\textwidth]{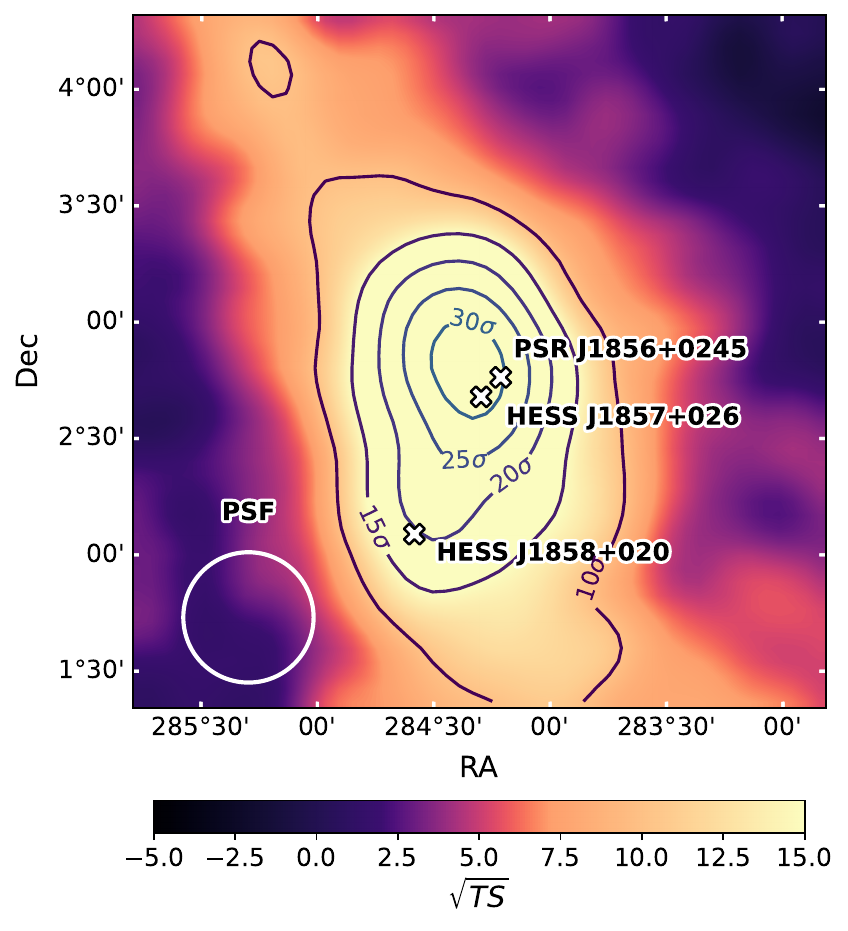}%
        \includegraphics[width=0.33\textwidth]{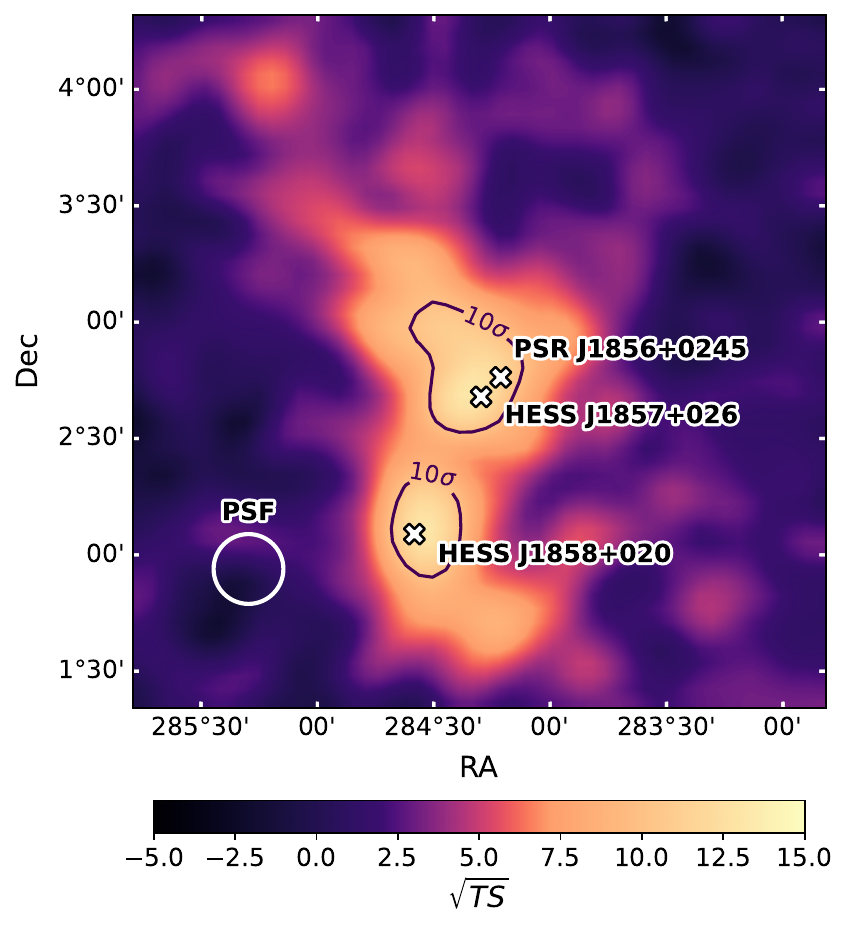}%
        \includegraphics[width=0.33\textwidth]{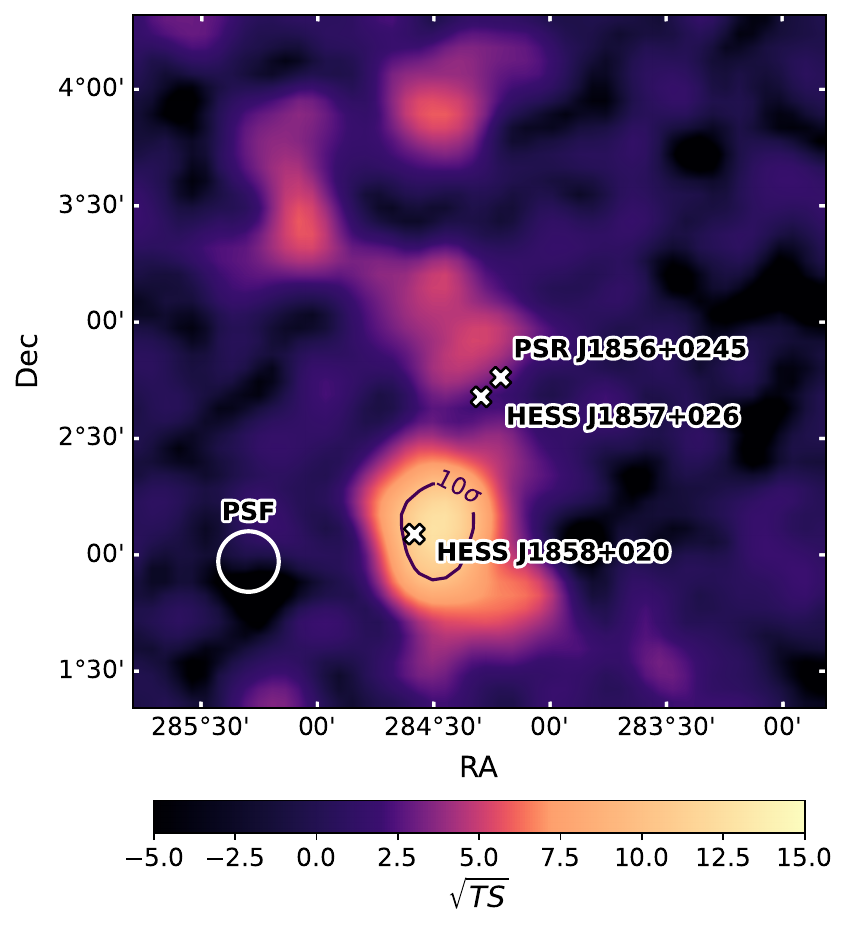}    
    \caption{
    HAWC significance maps \rev{in J2000 equatorial degrees} of the HESS~J1857+026 region in three energy ranges: 1 to 10 TeV (\textit{left}), 10 TeV to 31.6 TeV (\textit{middle}) and 31.6 to 316 TeV (\textit{right}). Emission from HESS~J1857+026 cuts off above an energy of 31.6 TeV. \hrev{The circles shown at the bottom left corner of the maps encompass the 68\% containment of the point spread function (PSF) obtained from the sum of individual bin PSF histograms for the corresponding energy range, weighted by the excess$^2$/bkg counts per bin for the given energy range.} \rev{The 10, 15, 20, 25, and 30$\,\sigma$ significance contours are also shown.} \hrev{Labels mark the positions of source associations in the HESS J1857+026 region.}
    }
    \label{fig:hawc_energy_range}
\end{figure*}
% uses the fit to the lateral distribution function to measure the charge density 40-m from the shower core, along with the zenith angle of the air shower, and estimates the energy of the primary gamma-ray.

A forward-folding method is performed to fit the spectral and spatial shape of the sources in the ROI using a maximum likelihood technique, maximizing the TS so that the input parameters have the highest likelihood of providing a good description of the observed data, following Eqn~\ref{equation:ts}. %The TS is defined as 
%\begin{equation}
%TS \equiv 2 \ln\frac{\mathcal{L}(H_1)}{\mathcal{L}(H_0)},
%\end{equation}
%where $\mathcal{L}$ is the likelihood function, $H_0$ is the background hypothesis, and $H_1$ is the signal plus background hypothesis, which depends on the spectral and spatial parameters assumed to describe the source. 
The ROI is a rectangular region of width $4^{\circ}$ in longitude and height of $20^{\circ}$ in latitude centered on the location of HESS~J1857+026 \hrev{for the correct estimation of diffuse background emission}. A multi-source fit pipeline \citep{alps_pipeline} adapted from the Fermi--LAT methodology is carried out in the region \citep[e.g.,][]{ackermann2017} using \rev{the Multi-Mission Maximum Likelihood\footnote{\url{https://github.com/threeML/threeML}} (3ML) python framework \citep{vianello2015multimissionmaximumlikelihoodframework} with the HAWC Accelerated Likelihood\footnote{\url{https://github.com/threeML/hawc_hal}} (HAL) plugin} \citep{hal_proceedings}.
%The pipeline models data by adding sources sequentially and performing a $\Delta$TS comparison at each step. A source is accepted into the model if the $\Delta \text{TS}\geqslant 25$. When a source is introduced into the model, its position and spectral parameters are fitted, while the positions of all other sources are fixed. The source-addition stops when $\Delta \text{TS}\leqslant 25$. Sources within the region are tested for extension assuming a Gaussian spatial shape, starting with the brightest source. A source is accepted as extended if $\Delta \text{TS} \geqslant 16$ in the overall model. Otherwise, the source is left as a point source. After all sources are tested for extension, the overall model is refitted and all parameters for point sources with $\text{TS} \geqslant 25$ are allowed to float. Any point source with $\text{TS} \leqslant 25$ is removed and a refit of the overall model is performed. The last part of the pipeline tests for curvature for each source using a log-parabola (LP) instead of a power law in order of decreasing TS. The new source is accepted as part of the model if $\Delta \text{TS} \geqslant 16$, otherwise the source's spectrum is left as a PL and the test continues for the next source.
A new source is added to improve the global fit as a point source with a power-law spectrum, with its spatial and spectral parameters free. The location of other sources remains fixed. The source is accepted into the model if the $\Delta \text{TS} > 25$. 

In the next step, each source is tested for extension using a Gaussian spatial template. If the extension of a source improves the global model by $\Delta \text{TS} > 16$, the best-fit Gaussian size is adopted. Once the new sources are characterized, the new global model is refitted, allowing all point sources with $\text{TS} > 25$ to vary in location. \hrev{Point sources} with $\text{TS} < 25$ are removed from the model. A final step performs curvature tests on each source using a log-parabola (LP) spectrum,
\begin{equation}
    \frac{dN}{dE} = N_{0} \left(\frac{E}{E_b}\right)^{-\Gamma-\beta\log{E/E_b}}.
\end{equation}\label{eq:logp}
A spectrum is considered curved if $\Delta \text{TS} \geqslant 16$, otherwise the spectrum remains the best-fit power-law. 

\hrev{The Galactic diffuse emission (GDE) is estimated using the High-Energy Radiative Messengers (HERMES) framework\footnote{\url{https://github.com/cosmicrays/hermes}} \citep{HermesCode} which models Galactic radiative processes. For the HAWC energy range, we only consider \rev{$\gamma$-ray} emission from $\pi^{0}$ and IC interactions \citep{HermesCode}. The diffuse IC is modeled from CR electrons and positrons \rev{upscattering} low-energy ambient photons from interstellar radiation fields (ISRF), i.e., UV, optical, infrared and the cosmic microwave background (CMB). The diffuse emissivities of IC are estimated from \cite{porter_inverse_2008} and the electron/positron CR spectrum is estimated using GALPROP \citep{strong2009galpropcosmicraypropagationcode,moskalenko2019icrc} where both are assumed to be isotropic. The $\pi^{0}$ emissivity is estimated from CR proton and \rev{helium} interactions with \rev{interstellar gas nuclei, assumed to be a mixture of neutral (HI) and molecular (H$_2$) hydrogen} \citep{acero_development_2016,HermesCode}. After estimating the predicted gamma-ray emission, an intensity map is generated as a 3D template cube (\rev{in equatorial coordinates} and binned in energy from 0.1 to 1000 TeV) and is added with dimensions that match the ROI size.}

The final model for the ROI consists of 4 extended sources (see Table~\ref{tab:hawc-res}): \hrev{HAWC} J1854+0120 (R.A.=$283^{\circ}.58\pm0.10$, Dec.=$1^{\circ}.34\pm0.12$), \hrev{HAWC} J1857+0200 (R.A.=$284^{\circ}.47\pm0.02$, Dec.=$2^{\circ}.00\pm0.02$), \hrev{HAWC} J1857+0247 (R.A.=$284^{\circ}.34\pm0.01$ Dec.=$2^{\circ}.80\pm0.02$), and \hrev{HAWC} J1858+0344 (R.A.=$284^{\circ}.74\pm0.05$, Dec.= $3^{\circ}.73\pm0.05$). \hrev{The sources HAWC J1854+0120 and HAWC J1858+0344 coincide with two LHAASO sources, 1LHAASO J1852+0050u* and 1LHAASO J1858+0330, respectively. However, in the 1LHAASO catalog, the emission from these sources suffers from background contamination \citep{cao2023first}.} 

The sources \hrev{HAWC} J1857+0247 and \hrev{HAWC} J1857+0200 are associated with HESS~J1857+026 and HESS~J1858+020 %\footnote{This source is also detected and modeled in the Fermi--LAT and VERITAS data. It is unlikely to be related to HESS~J1857+026, \hrev{see Section~\ref{sec:fresults} for more details.}}, 
respectively. \hrev{HAWC} J1857+0247 is the only source that shows evidence of curvature. The other sources assume a power-law spectrum. The spatial and spectral parameters for the final model are reported in Table~\ref{tab:hawc-res}. \rev{Significance maps of the final source model are presented in Figure~\ref{fig:hawc_energy_range}} \hrev{with cataloged source locations overlaid.}
 % Inaddition to modeling HESS~J1857+026 and the Galactic diffuse emission (GDE),  two more sources are included: HESS~J1858+020 
% and MAGIC~J1857.6+0297. %The Galactic diffuse emission (GDE) is also accounted for.
%We test a power law with an exponential cutoff (PLEC) spectrum compared to the power-law and log-parabola models for J1857+0247. For each spectral model (power-law, LP, or PLEC), we test two spatial models: a Gaussian and a diffusion model. %All spectral and spatial parameters are allowed to vary for the other sources reported from the multisource pipeline with the pivot energy fixed at 10\,TeV.
% HESS~J1858+020 is modeled as an extended Gaussian source with a log-parabola spectrum, while MAGIC~J1857.6+0297 is modeled as a point source with a power-law spectrum. For HESS J1857+026, we test three spectral shapes: a power law, a power law with an exponential energy cutoff, and a log-parabola. For each spectral shape, we test two spatial models: a Gaussian and a diffusion model. All spectral and spatial parameters are allowed to vary with the pivot energy fixed at 10\,TeV for all three sources. 
% \textcolor{red}{ }
% The GDE is modeled as a Gaussian distribution centered in the Galactic plane, and described by a power-law spectrum with index --2.75 \citep{gde1,gde2}; both the Gaussian width $\sigma$ and the normalization flux are left free in the fit.

\subsection{Morphology Sampling}
\hrev{To determine the morphological and spectral assumption that best describes the emission of HESS J1857+026, a model sampling is performed. The model samples over two spatial shapes: a symmetric Gaussian and a diffusion model (see Section~\ref{sec:pwn_physics}), and three spectral shapes: a power law, a log parabola and a power law with an exponential cutoff. All other sources assume the morphology and spectrum from the multi-source fit.} \frev{For non-nested models, the best-fit model is selected using the Bayesian information criterion \citep[BIC,][]{kassBayesFactors1995a, liddleInformationCriteriaAstrophysical2007} and the Akaike information criterion \citep[AIC,][]{bozdoganModelSelectionAkaikes1987}.} %Both the BIC and AIC help prevent the overfitting of a model by placing a penalty with the introduction of additional parameters. 
The BIC can be estimated as
\begin{equation}
    \text{BIC} = k \ln n - 2 \ln \mathcal{L}_{1}, \label{equation:bic}
\end{equation}
%where the penalty is placed using $k \ln n$, 
where $k$ is the number of free parameters in the model, $n$ is the number of independent observations, and $\ln \mathcal{L}_{1}$ is the maximized log-likelihood. The AIC is estimated by %The AIC also places a less strict penalty with $2 k$,
\begin{equation}
    \text{AIC} = 2k - 2 \ln \mathcal{L}_{1}.
\end{equation}
% We find that for the diffusion or Gaussian spatial model tested for HESS~J1857+026, the \hrev{gamma-ray} spectrum is best described by a power law with an exponential \rev{cutoff}
\hrev{The preferred model yields the lowest BIC and AIC values. We find that a power law with an exponential cutoff is strongly preferred for both the diffusion and Gaussian spatial models tested for HESS J1857+026:}
\begin{equation}
    \frac{dN}{dE} = K \left( \frac{E}{10 \ \rm{TeV}} \right)^{-\Gamma} \exp \left(- \frac{E}{E_{c}} \right).
    \label{equation:hawc_plec}
\end{equation}
\hrev{In the case for the diffusion spatial template,} $K=5.4(_{-1.4}^{+2.0})_{\text{stat}}(_{-2.3}^{+3.1})_{\text{sys}}\times 10^{-14}\;\rm{TeV}^{-1}\,\rm{cm}^{-2}\,\rm{s}^{-1}$, spectral index $\Gamma=2.11\pm 0.14_{\text{stat}} \pm 0.25_{\text{sys}}$, and cutoff energy $E_c = 14(_{-4}^{+5})_{\text{stat}}(_{-6}^{+7})_{\text{sys}}$ TeV, with an extension of $0.75 \pm 0.05_{\text{stat}} \pm 0.20_{\text{sys}}\,^\circ$. \hrev{The best-fit parameters for HAWC J1854+0120, HAWC J1857+0200, and HAWC J1858+0344 are consistent within uncertainty between the diffusion and Gaussian scenarios for HAWC J1857+0247.} The difference between the Gaussian and diffusion spatial models is $\Delta$BIC $= 17.8$ and $\Delta$AIC $=17.4$. This suggests that there is a strong preference for the diffusion model under a magnetic field of 1\,$\mu$G. However, the statistical preference is for the Gaussian if the magnetic field in the diffusion model is 5\,$\mu$G. \hrev{The best-fit parameters are provided in Table \ref{tab:hawc-res-5ug} for comparison.}

\begin{figure*}
\centering
\includegraphics[width=0.38\linewidth]{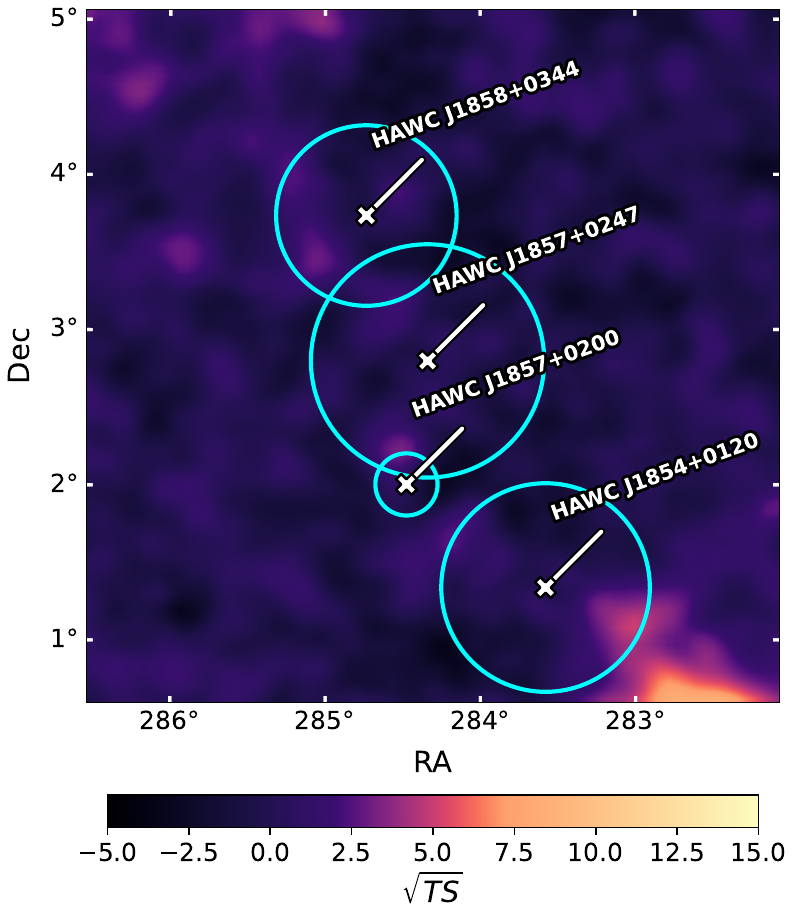}
\centering
\includegraphics[width=0.58\linewidth]{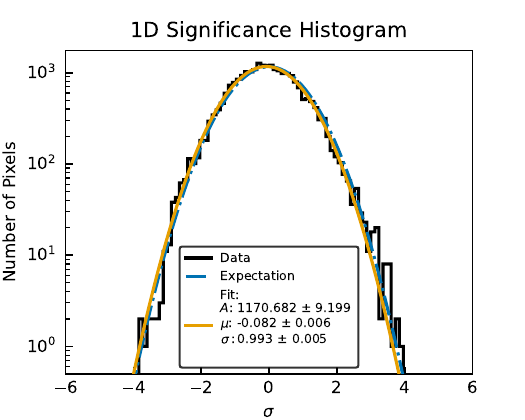}
\caption{{\it Left:} HAWC residual significance map in equatorial coordinates. The locations and extensions of sources comprising the final source model are displayed. {\it Right:} HAWC residual map projected into a 1D histogram. The Gaussian fit values are shown where $A$ is the normalization, $\mu$ is the mean value, and $\sigma$ the variance.
% The excess $\sim 4\sigma$ shown in the histogram is probably from the southeastern region of the ROI (see left panel) and likely from Galactic diffuse emission.
}\label{fig:hwc_residuals}
\end{figure*}

\rev{Given the strong dependence on the magnetic field strength for electron energy losses in the diffusion spatial template}, we provide the results of both the Gaussian and diffusion spatial models in Table~\ref{tab:hawc-res}.
%For completenes, we provide the details of the other model in Appendix X.}The best-fit spectral model to describe HESS~J1857+026 is a log-parabola,
% \begin{equation}
%     \frac{dN}{dE}=K\left(\frac{E}{10\,\rm{TeV}}\right)^{-\alpha -\beta\ln(E/10\,\rm{TeV})},
% \end{equation}
%The log-parabola spectral parameters for HESS~J1858+020 are $K=(1.0\pm 0.1)\times 10^{-14}\;\rm{TeV}^{-1}\,\rm{cm}^{-2}\,\rm{s}^{-1}$, spectral index $\alpha=-2.35\pm 0.09$, and $\beta = 0.20 \pm 0.06$ with a Gaussian extension of $0.17\pm0.02\,^\circ$. The power-law spectral parameters for the point source MAGIC~J1857.6+0297 are $K=(2.9\pm 0.1)\times 10^{-16}\;\rm{TeV}^{-1}\,\rm{cm}^{-2}\,\rm{s}^{-1}$ and spectral index $\alpha=-2.7\pm 0.2$. Finally, the GDE is fit using a power-law spectrum with $K=(6.0\pm 0.3)\times 10^{-11}\;\rm{TeV}^{-1}\,\rm{cm}^{-2}\,\rm{s}^{-1}\,{sr}^{-1}$ with a {\color{red} Gaussian} extension $0.63\pm0.04\,^\circ$. See Table~\ref{tab:hawc-res} for a summary of the results.
Figure~\ref{fig:hwc_residuals} shows the residual significance map (left panel) and histogram (right panel) of the ROI, highlighting that no additional excess emission $\gtrsim 5 \sigma$ within the ROI persists for the best-fit model \rev{assuming the diffusion spatial template for HESS~J1857+026 for $B = 1\,\mu$G}.
% Figure~\ref{fig:hwc_residuals} shows the residual significance map of the ROI and the residual histograms.

\begin{table}
\scalebox{0.9}{
\hspace{-1.75cm}
\begin{tabular}{l c c}
\hline
\hline
Parameter  & \multicolumn{2}{c}{Best-fit value} \\
& (Gaussian Model) & (Diffusion Model) \\
\hline
& \multicolumn{2}{c}{HAWC J1854+0120} \\
\hline
         % $\sigma$ & \multicolumn{2}{c}{$0.^{\circ}67\pm0.06_{\text{stat}} \pm 0.07_{\text{syst}}$} \\
         $\sigma$ & $0^{\circ}.73\pm0.07_{\text{stat}}(^{+0.30}_{-0.35})_{\text{syst}}$ & $0.^{\circ}67\pm0.06_{\text{stat}} \pm 0.30_{\text{syst}}$  \\
         $K$ & $18.8(^{+4.0}_{-3.1})_{\text{stat}}(^{+20}_{-15})_{\text{syst}}$ & $16(_{-2.2}^{+2.4})_{\text{stat}}(_{-11}^{+18})_{\text{syst}}$\\
         $\Gamma$ & $2.73\pm0.03_{\text{stat}}\pm 0.10_{\text{syst}}$ & $2.75\pm0.03_{\text{stat}} \pm 0.12_{\text{syst}}$ \\
\hline
& \multicolumn{2}{c}{HAWC J1857+0200 (HESS J1858+020)} \\
         \hline
        $\sigma$ & $0.^{\circ}20 \pm 0.01_{\text{stat}} \pm 0.04_{\text{syst}}$ & $0.^{\circ}20\pm0.01_{\text{stat}} \pm 0.03_{\text{syst}}$ \\
         $K$ & $9.4 \pm 0.6_{\text{stat}} \pm 1.8_{\text{syst}}$ & $8.9\pm0.6_{\text{stat}} \pm 1.7_{\text{syst}}$ \\
         $\Gamma$ &  $2.47 \pm 0.03_{\text{stat}} \pm 0.08_{\text{syst}}$ & $2.47\pm0.03_{\text{stat}} \pm 0.09_{\text{syst}}$ \\
\hline
& \multicolumn{2}{c}{HAWC J1857+0247 (HESS J1857+026)} \\
% & (Gaussian Model) & (Diffusion Model) \\
    \hline
         $\sigma$ & $0.^{\circ}26\pm (0.02)_{\rm stat}\pm (0.03)_{\rm syst}$ & $0.^{\circ}75\pm0.05_{\text{stat}}\pm0.20_{\text{syst}}$ \\
         $K$ & $43.8 (_{-5}^{+6})_{\text{stat}} (_{-28}^{+39})_{\text{syst}}$ & $54 (_{-14}^{+20})_{\text{stat}}(_{-23}^{+31})_{\text{syst}}$ \\
         $\Gamma$ & $2.18\pm 0.07_{\text{stat}}\pm0.40_{\text{syst}}$ & $2.11\pm0.14_{\text{stat}}\pm0.25_{\text{syst}}$\\
         $E_c$ & $13.5(_{-1.7}^{+1.9})_{\text{stat}}(_{-9}^{+13})_{\text{syst}}$ & $14(_{-4}^{+5})_{\text{stat}}(_{-6}^{+7})_{\text{syst}}$ \\
         \hline
     &   \multicolumn{2}{c}{HAWC J1858+0344} \\
\hline
         $\sigma$ & $0^{\circ}.61\pm0.06_{\text{stat}}(^{+0.21}_{-0.27})_{\text{syst}}$ & $0.^{\circ}58\pm0.08_{\text{stat}} \pm 0.33_{\text{stat}} $ \\
         $K$ & $15.1(^{+3.0}_{-2.5})_{\text{stat}}(^{+10}_{-12})_{\text{syst}}$ & $12.6(_{-2.4}^{+3.0})_{\text{stat}} (_{-9}^{+14})_{\text{syst}} $\\
         $\Gamma$ & $2.68\pm0.03_{\text{stat}}\pm 0.08_{\text{syst}}$ & $2.69\pm0.04_{\text{stat}} \pm 0.09_{\text{syst}} $\\
        \hline
        $K_{\text{GDE}}$ & $1.47\pm0.33_{\text{stat}}(^{+1.7}_{-1.3})_{\text{syst}}$ &  $1.53\pm0.25_{\text{stat}}\pm1.18_{\text{syst}} $ \\
\hline
\hline
\end{tabular}}
\caption{HAWC best-fit results in the 0.7--37\,TeV energy range. We report the values for both the diffusion ($B=$1\,$\mu$G) and Gaussian models for HESS~J1857+026. The normalization flux values $K$ have units $\times 10^{-15} \ \rm{TeV}^{-1}\,\rm{cm}^{-2}\,\rm{s}^{-1}$. $E_c$ is in TeV. \rev{$\sigma$ corresponds to the Gaussian} \hrev{width}.
}
% $K_{\rm GDE}$ has units $\rm{TeV}^{-1}\,\rm{cm}^{-2}\,\rm{s}^{-1}\,{\rm sr}^{-1}$.
\label{tab:hawc-res}
\end{table}

%\begin{figure*}
%\gridline{
%\fig{example-image-a}{0.33\textwidth}{(a)}
%\fig{example-image-b}{0.33\textwidth}{(b)}
%\fig{example-image-c}{0.33\textwidth}{(c)}
%}
%\caption{
%...
%}
%\label{fig:hwc_flux_map}
%\end{figure*}

%The HAWC data $\gamma$-ray significance maps in the energy range 3.16--100\,TeV are shown in Figure~\ref{fig:hwc_flux_map}.
%{\color{red} These HAWC maps show a consistent trend of correlation between the TeV emission and the pulsar PSR J1907+0602, where the $\gamma$-ray emission is seen to be more concentrated around the pulsar at higher energies. For MGRO J1908+06, the best-fit normalization flux is $K=(6.53\pm 0.26)\times 10^{-14}\;\rm{TeV}^{-1}\,\rm{cm}^{-2}\,\rm{s}^{-1}$, spectral index $\alpha=2.395\pm 0.025$, curvature $\beta = 0.157 \pm 0.014$ and extension $\sigma = 0.^{\circ}488\pm 0.^{\circ}014$. The list of best-fit parameters, including the other three sources and systematic uncertainties, is shown in Table \ref{tab:hawc-res}. The resulting $TS$ values for MGRO J1908+06, SS433E, SS433W and the GDE are 2050.05, 44.0, 63.7 and 276.9, respectively. Compared to the previous results reported in \cite{albert2022hawc}, in this work we report a higher TS value of $\Delta TS = 120$, and the best-fit normalization flux varies $\sim 30\%$ within statistical errors.

%\section{HAWC Systematic Uncertainties}
The detector performance and simulations produce a series of systematic uncertainties that are described in detail in \cite{hawc2017} and \cite{hawc2019}. The spectral and spatial parameters with positive and negative shifts are added in quadrature to account for the upward and downward uncertainties, respectively. These uncertainties are included in Table \ref{tab:hawc-res}.

\section{Multiwavelength Modeling}\label{sec:mw}
A favored scenario developed in \citet{guo2023} is one where SrcA and 4FGL~J1857.9+0313c have a hadronic origin such as \rev{an SNR interacting with molecular material}, while SrcT is more likely to be the PWN associated with PSR~J1856+0245. The hadronic model prediction for SrcA and 4FGL~J1857.9+0313c is estimated for distances $d  = 3.7\,$kpc and $d  = 6.3\,$kpc based on molecular material in those regions. %Similarly, 4FGL~J1857.9+0313c may be spatially coincident with molecular material at either close to PSR~J1856+0245, assuming 6.3\,kpc, or to molecular clumps found at 3.7\,kpc. 
It is possible that both SrcA and 4FGL~J1857.9+0313c may be hadronic counterparts to an unforeseen SNR shell associated with HESS~J1857+026. It is also possible that one or both sources are unrelated and are only positionally coincident to the TeV emission. No identified counterpart can be found for SrcA nor 4FGL~J1857.9+0313c \citep{guo2023}. In the following section, we consider both hadronic and lepto-hadronic particle populations to determine the more likely origin for HESS~J1857+026.

\subsection{Time-independent radiative models with NAIMA}\label{sec:naima}
We model the broadband emission exploring two basic scenarios: a hadronic (SNR) population and a lepto-hadronic population (PWN/SNR). The multiwavelength SED is fitted using the NAIMA python package \citep{naima} considering the \hrev{new Fermi--LAT, VERITAS, and HAWC data as well as available HESS \citep{hessgps2018} and MAGIC  \citep{magic2014} data}. In all models, we adopt the distance $d_{PSR} = 5.5$\,kpc as in \citet{petriella2021}. %, and $\beta$ is the spectral softening factor due to radiative cooling. {\color{red}. %For the single leptonic population, the best-fit parameters are: $A(d_{\mathrm{PSR}}) = (9.8 \pm 2.0)\times10^{45} \times (d_{\mathrm{PSR}}/\mathrm{kpc})^{2}\ \mathrm{TeV}^{-1}$, $E_{\mathrm{break}} = 9.2 \pm 1.4 \ \mathrm{TeV}$, and $\beta = 2.1 \pm 0.3$.}

\begin{figure*}
\begin{minipage}[b]{0.5\textwidth}
\centering 
\includegraphics[width=1.0\linewidth]{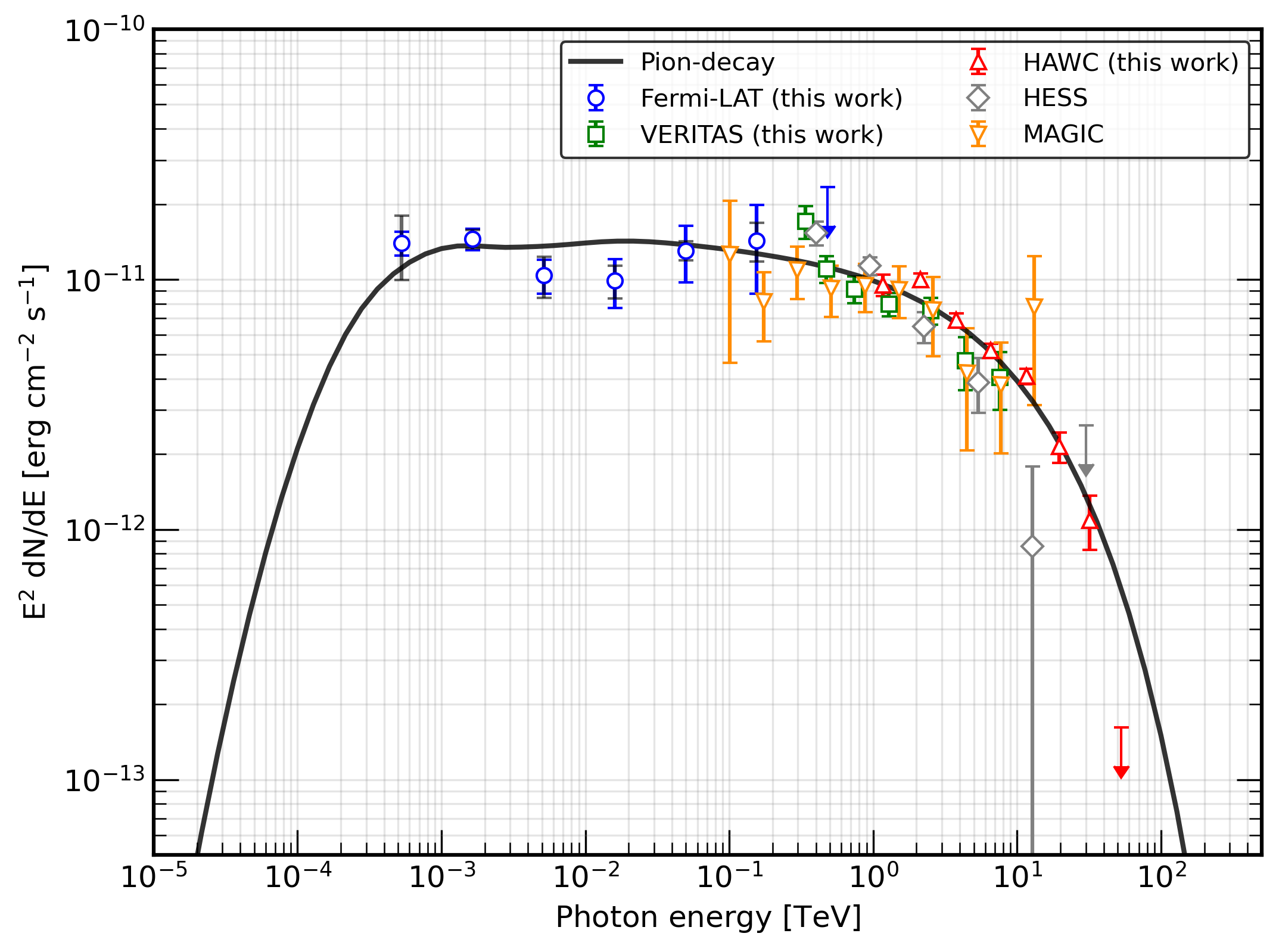}
\end{minipage}
\begin{minipage}[b]{0.5\textwidth}
\centering 
\includegraphics[width=1.0\linewidth]{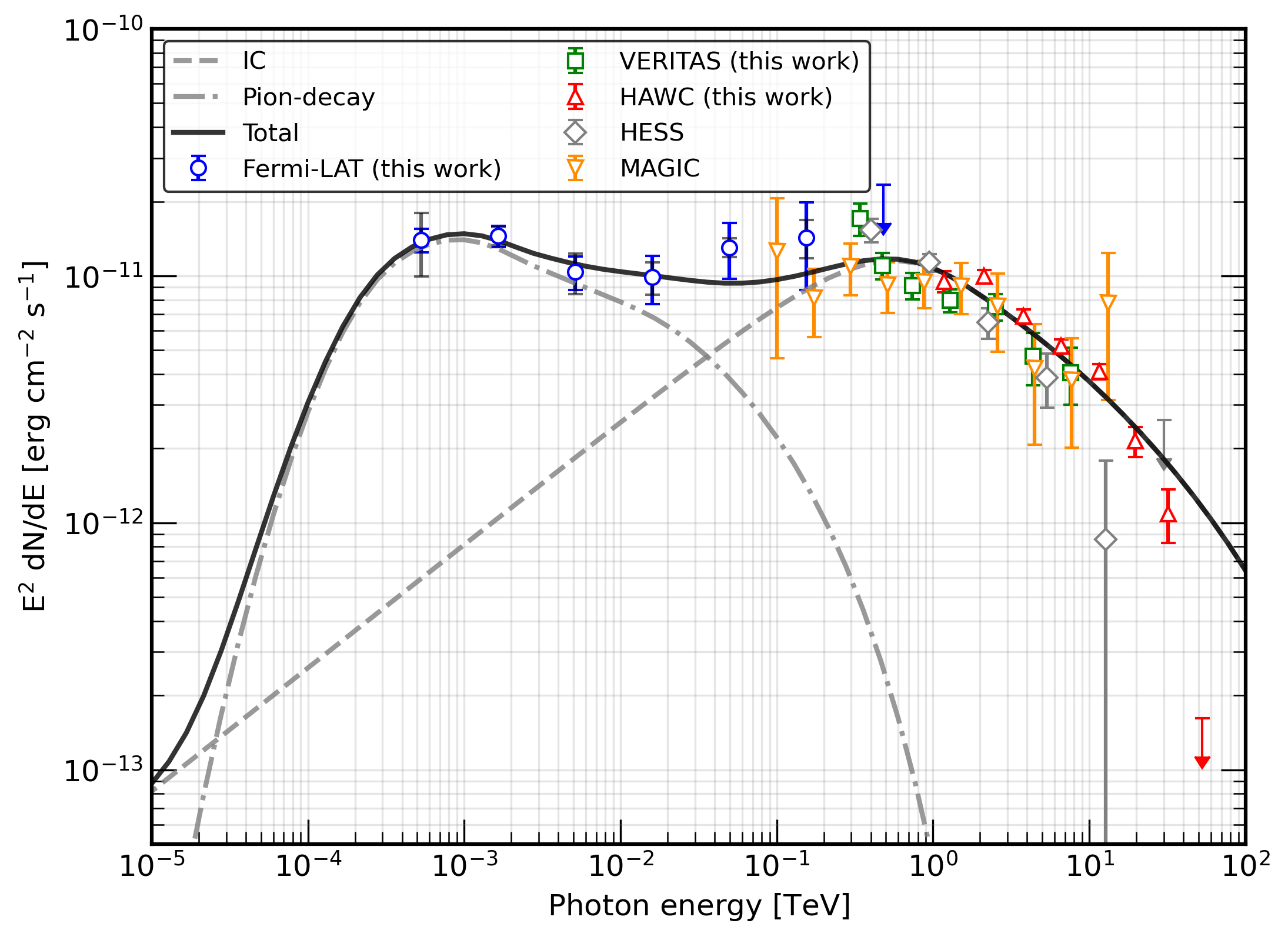}
\end{minipage}
\caption{\rev{{\it Left:}} Hadronic scenario. \rev{{\it Right:}} Lepto-hadronic scenario. {\it Both panels:} The results of the time-independent NAIMA SED fit. In blue are the Fermi--LAT flux data points for $E > 300\,$MeV (this work), in green are VERITAS flux points (this work), and in red are HAWC flux points (this work). We also include TeV data from HESS \citep{hessgps2018} and MAGIC \citep{magic2014} in gray and orange, respectively. The Fermi--LAT systematics (black) are those of \citet{eagle_2022}. }
\label{fig:MW_SED_fitting}
\end{figure*}

%A hadronic pion decay scenario is also explored, since the SNR shell identification remains unclear in broadband observations in addition to dense surroundings that may enable an energetic interaction between the SNR and ambient media. %We derive an ambient density estimate from the neutral hydrogen cloud distribution in the direction of HESS~J1857+026, following \citet{petriella2021}, see Figure~\ref{fig:cloud_map}. We find an estimate that is $\rho \sim xx$\,cm$^{-3}$, and which agrees with the estimate reported in \citet{petriella2021}. 
In the hadronic scenario, \rev{the $\gamma$-ray emission is attributed to high-energy protons interacting with ambient gas leading to pion decay. The ambient density estimated in \citet{petriella2021}, 22\,cm$^{-3}$, is used for the pre-shocked particle density $n_0$}. A hadronic scenario becomes plausible at the SNR forward shock \rev{for a strong, unmodified shock with a compression ratio of 4. In this case,} \rev{the post-shocked particle density is at least 4 times the pre-shocked value,} $n_h \gtrsim 4 n_0$. %\citep[see e.g.,][and references therein]{castro_2013}. 
We fix $n_h$ to $88$\,cm$^{-3}$ in the hadronic model. \rev{The proton} distribution is constant in time and is assumed to be a \rev{power-law with an exponential cutoff (PLEC)} as a function of pulsar distance ($d_{PSR}$) and energy $E$,

\begin{equation}
    f_{p}(d_{\mathrm{PSR}},E) = A_p(d_{\text{PSR}}) \big(\frac{E}{E_0}\big)^{-\alpha_{p}} \exp{\big(-\frac{E}{E_{p,\text{cut}}}\big)}\label{equation:protons}
\end{equation}

\iffalse
\begin{equation}\label{eq:naima_p}
\begin{aligned}
&f(d_{\mathrm{PSR}},E)
\\ 
&=\begin{cases}
A(d_{\mathrm{PSR}})\left(\frac{E}{E_{0}}\right)^{-\alpha}
, & \text{if } E < E_{\mathrm{break}} \\
A(d_{\mathrm{PSR}})\left(\frac{E_{\mathrm{break}}}{E_{0}}\right)^{\beta}
\left(\frac{E}{E_{0}}\right)^{-\alpha-\beta}
, & \text{if } E > E_{\mathrm{break}}
\end{cases}
\end{aligned}
\end{equation}
\fi
where $E_{0}=1$ TeV is the reference energy, $A_p(d_{\mathrm{PSR}})$ is the distance-dependent number of \rev{protons} per unit energy, $E_{p,\mathrm{cut}}$ is the \rev{proton} cutoff energy, and $\alpha_{p}$ is the \rev{proton} spectral index. \rev{A PLEC for the hadronic model is motivated by the less efficient cooling of protons such that their cooling times are comparable to or exceed the age of the system. This results in the proton spectrum being governed by the shock acceleration efficiency that naturally leads to an exponential cutoff as the highest energy protons escape the shock \citep[e.g.,][]{ohira2010}.} %, while assuming the same particle distribution in Eq.~\eqref{eq:naima_p} for protons. 
The best-fit parameters are $W_p(d_{\mathrm{PSR}}) = (2.6 \pm 0.1)\times10^{49}\ \mathrm{erg}$, $E_{p,\mathrm{cut}} = 122.3^{+17}_{-14}\,\mathrm{TeV}$, $\alpha_{p} = 2.10^{+0.01}_{-0.01}$ where $W_p(d_{\mathrm{PSR}})$ is the total proton energy.%, and $\beta = 4.22^{+1.33}_{-1.25}$. %We note that the total proton energy and the post-shocked particle density $n_h$ are inversely proportional. If the $n_h$ value is larger than the predicted xx\,cm$^{-3}$, then the total proton energy will decrease by $\frac{n_{h,\text{true}}}{n_{h,\text{HI}}}$ and vice versa.

In the lepto-hadronic scenario, we consider that the GeV emission is primarily explained by pion decay while the TeV emission is attributed to IC scattering off of the cosmic microwave background (CMB). 
%The electrons and protons are considered to be accelerated in different regions such as a SNR forward shock for the protons and the PWN termination shock for electrons. 
%In the hadronic scenario, we propose that the emission originates from protons accelerated in an old supernova remnant that may no longer be visible, interacting with a molecular cloud. 
The protons are characterized \rev{again as a PLEC, while the}
%\begin{equation}
%    f^{p}(d_{\mathrm{PSR}},E) = %A(d_{\text{PSR}}) \big(\frac{E}{E_0}\big)^{-\alpha^{p}} \exp{\big(-\frac{E}{E_{\text{cut}}^{p}}\big)}\label{equation:protons}
%\end{equation}
%
%As protons gain more energy, the shock's ability to confine them weakens, allowing the highest-energy protons to escape. 
%Conversely, in the leptonic scenario, we attribute the TeV emission to high-energy electrons accelerated in a pulsar wind nebula (PWN) interacting with ambient photons, primarily those from the Cosmic Microwave Background (CMB). 
distribution of electrons is represented by the broken power-law (BPL) defined as
\begin{equation}
\begin{aligned}
&f_{e}(d_{\mathrm{PSR}},E) = A_e(d_{\mathrm{PSR}})
\\ 
&\begin{cases}
\left(\frac{E}{E_{0}}\right)^{-\alpha_{e}}
, & \text{if } E < E_{e,\mathrm{break}} \\
\left(\frac{E_{e,\mathrm{break}}}{E_{0}}\right)^{\beta_{e}}
\left(\frac{E}{E_{0}}\right)^{-\alpha_{e}-\beta_{e}}
, & \text{if } E > E_{e,\mathrm{break}}
\end{cases}
\end{aligned}\label{equation:electrons}
\end{equation}
where $E_{0}=1$ TeV is the reference energy. %and $E_{e,\text{cut}}$ is the \rev{electron} cutoff energy. %The definitions for the rest of the parameters are already defined in Equation \ref{eq:naima_p}.
%Moreover, 
\rev{Efficient cooling of the highest-energy electrons in synchrotron and IC radiation would lead to a break in the spectrum.
%\textcolor{blue}{\textbf{This logic now does not apply:} The cooling timescale for electrons is likely shorter than the age of the system which causes a steepening in the spectrum above a characteristic energy ($E_{e,\text{break}}$). The suppression of the acceleration efficiency of the highest-energy electrons can additionally lead to an exponential cutoff beyond the characteristic energy ($E_{e,\text{cut}}$).}
We also tested a broken power-law with an exponential cutoff for the electron distribution, but the broken power-law model is statistically more favorable according to the BIC criterion ($\Delta$BIC = 17).} % The choice of a ECBPL for the electrons \rev{over a simpler model such as a broken power-law (BPL)} is \rev{statistically motivated}.  %by electron cooling and a statistical preference for the ECBPL over the BPL in the lepto-hadronic case. 
\rev{The resulting lepto-hadronic model has parameters for the protons \rev{following Equation~\ref{equation:protons}} as $W_p(d_{\mathrm{PSR}}) = (1.9 \pm 0.4)\times10^{49}\ \mathrm{erg}$, $\alpha_p = 2.3 \pm 0.1$, assuming $E_{p,\mathrm{cut}} = 1\,\mathrm{TeV}$ and $n_h = 88$\,cm$^{-3}$. The electron parameters following Equation~\ref{equation:electrons} are $W_e(d_{\mathrm{PSR}}) = (4.6 \pm 0.9)\times10^{48}\ \mathrm{erg}$, $E_{e,\mathrm{break}} = 9.7^{+1.9}_{-1.6}\,\mathrm{TeV}$, $\beta_e = 1.55^{+0.07}_{-0.06}$, %$E_{e,\text{cut}} = 52.2^{+8.0}_{-8.1} \ \mathrm{TeV}$ 
assuming $\alpha_e = 2.0$} where $W_e(d_{\mathrm{PSR}})$ is the total electron energy. The best-fit hadronic and lepto-hadronic models are displayed in Figure~\ref{fig:MW_SED_fitting}.

%Assuming 10\% of the canonical SN explosion energy $10^{51}$\,erg is carried away in CRs, 
The purely hadronic model suggests that $\sim 3\%$ of the SN explosion energy is carried away in CRs for $E_{\text{SN}} = 10^{51}$\,erg, which is not unreasonable, \hrev{compared to the canonical expectation $\sim$10\%}. \citet{petriella2021} assumed a \hrev{relatively high} value \hrev{for the fraction of SN explosion energy transferred to CRs of} 30\% and found that the TeV flux \hrev{$F_\gamma(E > \text{1\,TeV})$} could be explained by the following relation and conditions:
\begin{equation}
F_\gamma(E > \text{1\,TeV}) = 1\times10^{-10} f_\gamma \theta E_{51} D_{\text{kpc}}^{-2} n_0
\label{equation:tevflux}
\end{equation}
assuming the \rev{flux coefficient is} $f_\gamma =0.19$, \rev{a value determined by the power-law proton spectral index value 2.3 \citep{torres2003}}, \rev{the fraction of $E_{SN} = 10^{51}$\,erg in CRs is} $\theta = 0.3$, $E_{51} = 1$, \hrev{the distance is $D_{\text{kpc}} = 5.5$}, and $n_0 = 22$\,cm$^{-3}$, resulting in a TeV flux \hrev{$F_\gamma(E > \text{1\,TeV}) \sim 4 \times 10^{-12}$\,ph cm$^{-2}$ s$^{-1}$}. The TeV flux for HESS~J1857+026, for comparison, is $F_\gamma (1 > \text{1\,TeV}) = 3.77 \pm 0.4 \times 10^{-12}$\,ph cm$^{-2}$ s$^{-1}$ \citep{hessgps2018}. \apjrev{If the fraction of CR energy is $\theta <0.1$}, the TeV flux becomes too low to explain HESS~J1857+026 solely from hadronic emission. \rev{The same is true if $n_0$ is lower than 22\,cm$^{-3}$. For the parameters of the purely hadronic model, $\theta = 0.03$ and the proton spectral index value is 2.1, corresponding to $f_\gamma = 0.9$, and results in a TeV flux from Eqn~\ref{equation:tevflux}, $F_\gamma(E > \text{1\,TeV}) \sim 2 \times 10^{-12}$\,ph cm$^{-2}$ s$^{-1}$.}

The lepto-hadronic model is also plausible and could be attributed to either a lepto-hadronic SNR or a hadronic SNR and a leptonic PWN. 
In the case of a lepto-hadronic SNR, the implied \rev{average} electron to proton ratio is \rev{$k_{ep} \sim 0.2$, 20 times higher than the canonical value based on estimates measured on Earth \citep[$k_{ep} \sim 0.01$, e.g.,][]{merten2017}}, \rev{though with an energy-dependence introduced by the different particle spectral shapes, becoming $\sim 0.1$ at 5\,GeV up to $\sim 0.3$ at 100\,GeV}.

In conclusion, the presented models cannot rule out a hadronic component for the $E < 10\,$GeV $\gamma$-rays, but a PWN leptonic contribution is likely the dominant contribution to the observed $\gamma$-ray emission \hrev{and is supported by the improved fit to the data of the lepto-hadronic model over the hadronic model shown in Figure~\ref{fig:MW_SED_fitting}. The presence of an energetic pulsar coincident with the GeV--TeV $\gamma$-ray emission observed to concentrate near the pulsar with increasing energy supports a PWN leptonic origin\rev{, as indicated by the reduction of Gaussian radius from Fermi's 0.38$^\circ$ to HAWC's 0.26$^\circ$, a difference nearly 3 times the combined statistical uncertainties.}. The age of the system implies an evolved PWN that may have a low magnetic field strength which would explain an energy-dependent morphology in the GeV--TeV band and the lack of a PWN X-ray counterpart. Finally, no SNR emission has been detected in any wavelength in the region of the $\gamma$-ray emission.} 
In the following section, we expand the leptonic broadband characterization for the source emission assuming it is dominated by the PWN and consider basic evolution of the radiative properties.

\subsection{Time-dependent radiative models considering PWN evolution}\label{sec:yosi}

%JDG: suggest the following changes to the text below
%In the previous section, we favor a leptonic population such as a PWN origin. %that modeling the non-thermal broadband SED suggests that it most likely originates from a leptonic population such as a PWN origin. %\citep[see e.g.,][]{gelfand2009dynamical, torres2014}.
%To determine if a leptonic PWN can explain the intrinsic properties of this system, 
We model the observed properties of the PWN powered by PSR~J1856+0245 assuming it is responsible for the detected $\gamma$-ray emission as it evolves inside \rev{an} SNR shell.
%The current best fits parameters are (with some comments attached):
%SN explosion energy = 3.1e51 ergs
%SN ejecta mass = 4.0 Solar Masses (a bit on the low side, and what is expected for a massive progenitor that generated powerful %winds before it exploded)
%ISM density = 0.44 cm^-3  (lower than average)
%Pulsar braking index = 2.96
%Spin-down timescale = 84 years (lower than most systems, but higher than Kes 75)
%True age = 21 years
%Initial E-dot =  3.3e41 erg/s
%Wind Magnetization eta_B = 6.2e-4 (within normal range)
%Minimum injected particle energy =  18 GeV
%Maximum injected particle energy =  2.0 PeV
%Break energy in injected particle spectrum = 0.3 TeV
%Low energy particle index = 1.55 (typical)
%High energy particle index = 3.0 (a bit softer than normal)
%This fit had a chi^2 of 37 for 13 degrees of freedom, with the chi^2 dominated by two of the Fermi point that fall below the model prediction in the gamma-ray band.  I haven't calculated the model predicted radio and X-ray fluxes yet, but by eye it predicts an extremely soft X-ray source
%Attached are the model output files for this best fit
\begin{table*}
\centering
\begin{tabular}{|c c c c c|}
\hline
\/Shorthand & Parameter & Best-Fit & Best-Fit & Units \\
\/& & ($E > 300\,$MeV) & ($E > 10\,$GeV) &  \\
\hline
\hline
\ \texttt{chi2} & $\chi^2$ of Spectral Energy Distribution & 27  & 12 & -- \\
\ \texttt{dof} & degrees of freedom of Spectral Energy Distribution & 14 & 16 & -- \\
\hline
\ \texttt{esn} & Initial Kinetic Energy of Supernova Ejecta & 5.3 & 3.7 & $10^{51}$ ergs \\
\hline
\ \texttt{mej} & Mass of Supernova Ejecta & 3.5 & 5.5 & Solar Masses \\
\hline
\ \texttt{nism} & Number Density of Surrounding ISM & 0.13  & 0.01 & cm$^{-3}$ \\
\hline
\ \texttt{brakind} & Pulsar Braking Index & 2.96 & 2.96 & - \\
\hline
\ \texttt{tau} & Pulsar Spin-down Timescale & 400 & 5000 & years \\
\hline
\ \texttt{age} & Age of System & 21 & 16.4 & kyrs \\
\hline
\ \texttt{e0} & Initial Spin-down Luminosity of Pulsar & 1.4 & 0.00085 & $10^{41}$ ergs s$^{-1}$ \\
\hline

\ \texttt{etag} & Fraction of Spin-down Luminosity lost as Radiation & $\equiv0$ & $\equiv0$ & - \\
\hline
\ \texttt{etab} & Magnetization of the Pulsar Wind & 3.0 $\times 10^{-3}$ & 1.75 $\times 10^{-3}$ & - \\
\hline
\ \texttt{emin} & Minimum Particle Energy in Pulsar Wind & 30 & 4.5 & GeV \\
\hline
\ \texttt{emax} & Maximum Particle Energy in Pulsar Wind & 3.4 & 0.86 & PeV \\
\hline
\ \texttt{ebreak} & Break Energy in Pulsar Wind & 0.125 & 2.4 & TeV \\
\hline
\ \texttt{p1} & Injection Index below the Break & 1.22 & 1.89 & -- \\
\ & (${dN}/{dE} \sim E^{-p1}$) & & & \\
\hline
\ \texttt{p2} & Injection Index above the Break & 2.95 & 3.09 & -- \\
\ & (${dN}/{dE} \sim E^{-p2}$) & & & \\
\hline
%\ \texttt{ictemp} & Temperature of each Background Photon Field & xx & $10^{6}$ K \\
%\hline
%\ \texttt{icnorm} & Log Normalization of each Background Photon Field & xx & - \\
%\hline
\ \texttt{kpsr} & Log Normalization of Direct $\gamma$-ray Emission from the Pulsar & $\equiv0$ & $\equiv0$& -- \\
\hline
\ \texttt{gpsr} & Photon Index of the $\gamma$-rays Produced Directly by the Pulsar & $\equiv0$ & $\equiv0$& -- \\
\hline
\ \texttt{ecut} & Cutoff Energy from the Power Law of Pulsar Contribution & $\equiv0$ & $\equiv0$ &  GeV \\
\hline
\end{tabular}
\caption{Summary of the input parameters for the evolutionary system and their best fit values considering a PWN origin to the observed $\gamma$-ray emission.}\label{tab:inputoutputgelfand}
\end{table*}
We use the dynamical and radiative properties of a PWN predicted by an evolutionary model, similar to what is described by \citet{gelfand2009dynamical}, to identify the combination of neutron star, pulsar wind, supernova explosion, and ISM properties that can best reproduce what is observed. The model is developed using a Markov chain Monte Carlo (MCMC) fitting procedure \citep[see, e.g.,][for details]{gelfand2015} to find the combination of free parameters that can best represent the observations. The observed sizes of the SNR and PWN together with the $\gamma$-ray data are used to calculate the final broadband model at an age, $t_{age}$. The PWN angular radius range is 0.1 to 0.4$\,^\circ$, in accord with the extended $\gamma$-ray observations. The SNR angular radius range is 0.45 to 0.77$\,^\circ$ based on the size of the HI cavity. The predicted dynamical and radiative properties of the PWN that correspond to the best representation of the broadband data are listed in Table \ref{tab:inputoutputgelfand}. The parameters \texttt{etag, kpsr, gpsr, ecut}, and pulsar velocity are fixed to zero. The distance is fixed to 5\,kpc.

\begin{figure*}
\begin{minipage}[b]{0.5\textwidth}
\centering 
\includegraphics[width=1.0\linewidth]{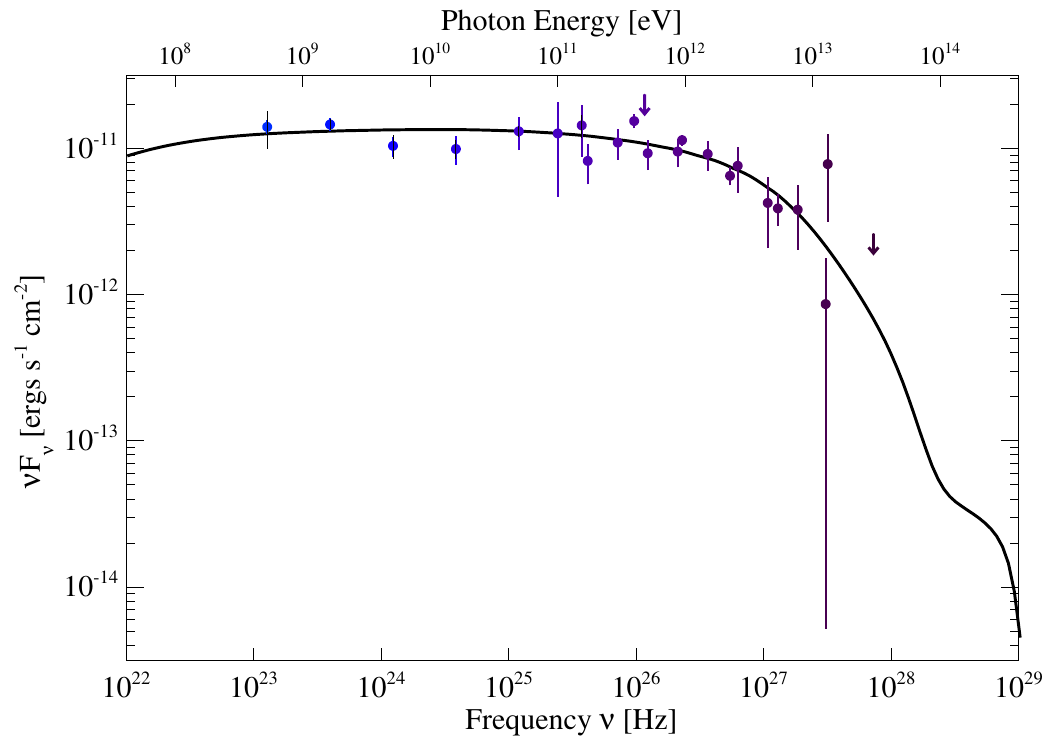}
\end{minipage}
\begin{minipage}[b]{0.5\textwidth}
\centering 
\includegraphics[width=1.0\linewidth]{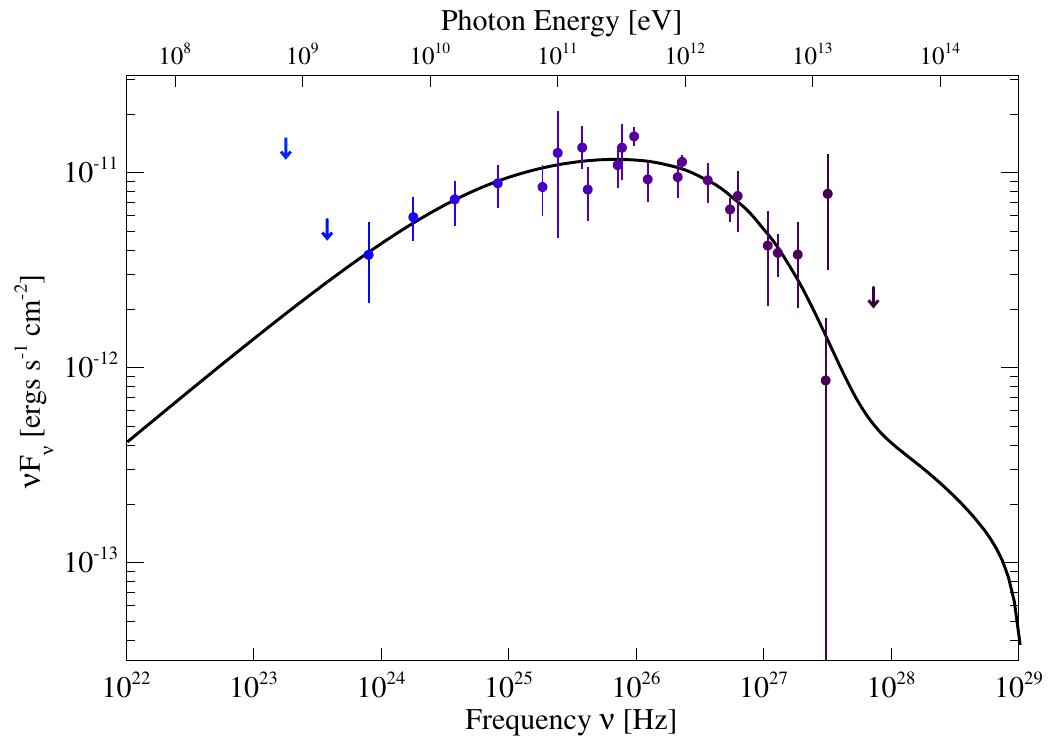}
\end{minipage}
\caption{{\it Left}: The best-fit SED obtained through the evolutionary model method described in Section~\ref{sec:yosi}. The colored points \rev{(color is proportional to photon energy)} represent the values of observed data that the model used as comparison points for fitting: the Fermi--LAT (blue, this work) and HESS and MAGIC \citep[purple,][]{hessgps2018,magic2014}. {\it Right}: The best-fit SED obtained through the evolutionary model using $E > 10\,$GeV Fermi--LAT data from \citet{guo2023}.}\label{fig:gelfand_sed}
\end{figure*} 

The analysis performed here is similar to what has previously been reported for G54.1+0.3 \citep{gelfand2015}, Kes~75 \citep{gotthelf2021,straal2022}, HESS~J1640--465 \citep{mares2021,moaz2023}, and B0453--685 \citep{eagle2023}. \rev{A pulsar with} characteristic age $t_{ch}$, \rev{defined to be $t_{\rm ch} \equiv \frac{P}{\dot{P}}$ where $P$ is the period of the associated pulsar and $\dot{P}$ is its period derivative,} \citep[see][]{pacini1973,gaensler2006},  \rev{has an actual age} $t_{age}$ \rev{of}
\begin{equation}
  t_{\rm age} = \frac{2t_{\rm ch}}{p-1} - \tau_{\rm sd},
\end{equation}
\rev{where $\tau_{\rm sd}$, often referred to as the spin-down timescale, is the birth characteristic age of the pulsar. The}
spin-down luminosity $\dot{E}$ \rev{evolves as}
\begin{equation}
  \dot{E}(t) = \dot{E_0}\big(1 + \frac{t}{\tau_{sd}}\big)^{-\frac{p+1}{p-1}}
\end{equation}
and is chosen for a braking index $p$, initial spin-down luminosity $\dot{E_0}$, and $\tau_{sd}$ to best reproduce the pulsar's characteristic age and current spin-down luminosity. A fraction $\eta_\gamma$ of this luminosity is converted to $\gamma$-ray emission from the neutron star's magnetosphere, \rev{while} the rest $(1-\eta_\gamma)$ is injected into the PWN in the form of a magnetized, highly relativistic outflow, i.e., the pulsar wind. The pulsar wind enters the PWN at the termination shock, where the rate of magnetic energy $\dot{E}_B$ and particle energy $\dot{E}_P$ injected into the PWN is expressed as:
\begin{eqnarray}\label{eqn:edotb}
\dot{E}_B(t) & \equiv & \eta_{\rm B}\dot{E}(t) \\
\dot{E}_P(t) & \equiv & \eta_{\rm P}\dot{E}(t)
\end{eqnarray}
where $\eta_B$ is the magnetization of the wind and defined to be the fraction of the pulsar's spin-down luminosity injected into the PWN as magnetic fields and $\eta_P$ is the fraction of spin-down luminosity injected into the PWN as particles. We assume the PWN ICS emission results from leptons scattering off the CMB as in Section~\ref{sec:naima}. %, however the total particle energy and the properties of the background photon fields cannot be independently determined. Since the evolutionary model accounts for the decline in total particle energy from the adiabatic losses of early PWN evolution and the increase of synchrotron losses at later times from compression, where both likely have a significant effect on the oldest particles, a second photon field is hence required. The second, ambient photon field is defined by temperature $T_{IC}$ and normalization $K_{IC}$, such that the energy density of the photon field $u_{IC}$ is
%\begin{equation}
%  u_{IC} = K_{IC}a_{BB}T^4_{IC}
%\end{equation}
%where $a_{BB} = 7.5657 \times 10^{-15}$\,erg cm$^{-3}$ K$^{-4}$. 

\rev{Past studies of PWNe (e.g., \citealt{bucciantini11, torres2014}) have found that reproducing the broadband SED requires the spectrum of particles injected at the termination shock be described by a broken power-law.  A theoretical motivation for this choice comes from recent particle-in-cell simulations, which find that both the standard Fermi acceleration mechanism and magnetic reconnection are expected to accelerate particles at the termination shock (e.g., \citealt{sironi11, cerutti20}).} We \rev{therefore} assume the particle injection spectrum at the termination shock \rev{in this system} is \rev{also} well-described by a broken power-law distribution:
\begin{equation}
\frac{d\dot{N}_{e^\pm}(E)}{dE} =
\begin{cases}
 \dot{N}_{break} \big(\frac{E}{E_{break}}\big)^{-p_1} & E_{min} < E < E_{break} \\
 \dot{N}_{break} \big(\frac{E}{E_{break}}\big)^{-p_2} & E_{break} < E < E_{max} \\
\end{cases}
\end{equation}
where $\dot{N}_{e^\pm}$ is the rate that electrons and positrons are injected into the PWN, and $\dot{N}_{break}$ is calculated using
\begin{equation}
  \eta_P\dot{E} = \int_{E_{min}}^{E_{max}} E \frac{d\dot{N}(E)}{dE} dE.
\end{equation}
\rev{Following past work on similar systems (e.g., \citealt{gelfand2015}), we fit all  quantities related to the particle injection spectrum ($E_{\rm min}$, $E_{\rm break}$, $E_{\rm max}$, $p_1$, and $p_2$) and assume they are constant with time.}
We show the spectral energy distribution for HESS~J1857+026 that can reasonably reproduce the observed spectrum in Figure~\ref{fig:gelfand_sed}\rev{, with} the best-fit parameters in Table~\ref{tab:inputoutputgelfand}. Since we cannot rule out an unrelated low-energy ($E<10\,$GeV) component, we provide two resulting models: one using the Fermi--LAT data presented in Section~\ref{sec:fermi_analysis} and the second using the $E > 10$\,GeV data from \citet{guo2023}. If a PWN dominates $>$300\,MeV, the true age estimate is 21\,kyr \rev{with a magnetic field value for the PWN $B_{PWN} \sim 1.6\,\mu$G}. In the $>10\,$GeV case, the true age estimate is \apjrev{16\,kyr} \rev{and $B_{PWN} \sim 0.4\,\mu$G}. \rev{Both age} values are within the characteristic age estimate for PSR~J1856+0245. We also note that the ISM density in either model is very low, which implies that hadronic emission from the SNR is unlikely. %A pulsar contribution is plausible. 
\apjrev{The model prediction presented here is similar to the recent work of \citet{gong2026}, which assumes a distance of 6.3\,kpc and finds a magnetic field strength $2.6\,\mu$G, a system age $13\,$kyr, and particle indices before and after the break 1.4 and 3.2.}

% In the following section we evaluate and compare the model predictions to the observational constraints of the system. The results of the fits using data from Fermi--LAT, VERITAS, HAWC, and LHAASO data is shown in Figures \ref{fig:MW_SED_fitting} and \ref{fig:gelfand_sed}.

\section{Diffusion Properties of PWN HESS~J1857+026}\label{sec:pwn_physics}

\begin{figure*}[t!]
\begin{minipage}[b]{0.33\textwidth}
\centering 
\includegraphics[width=1.0\linewidth]{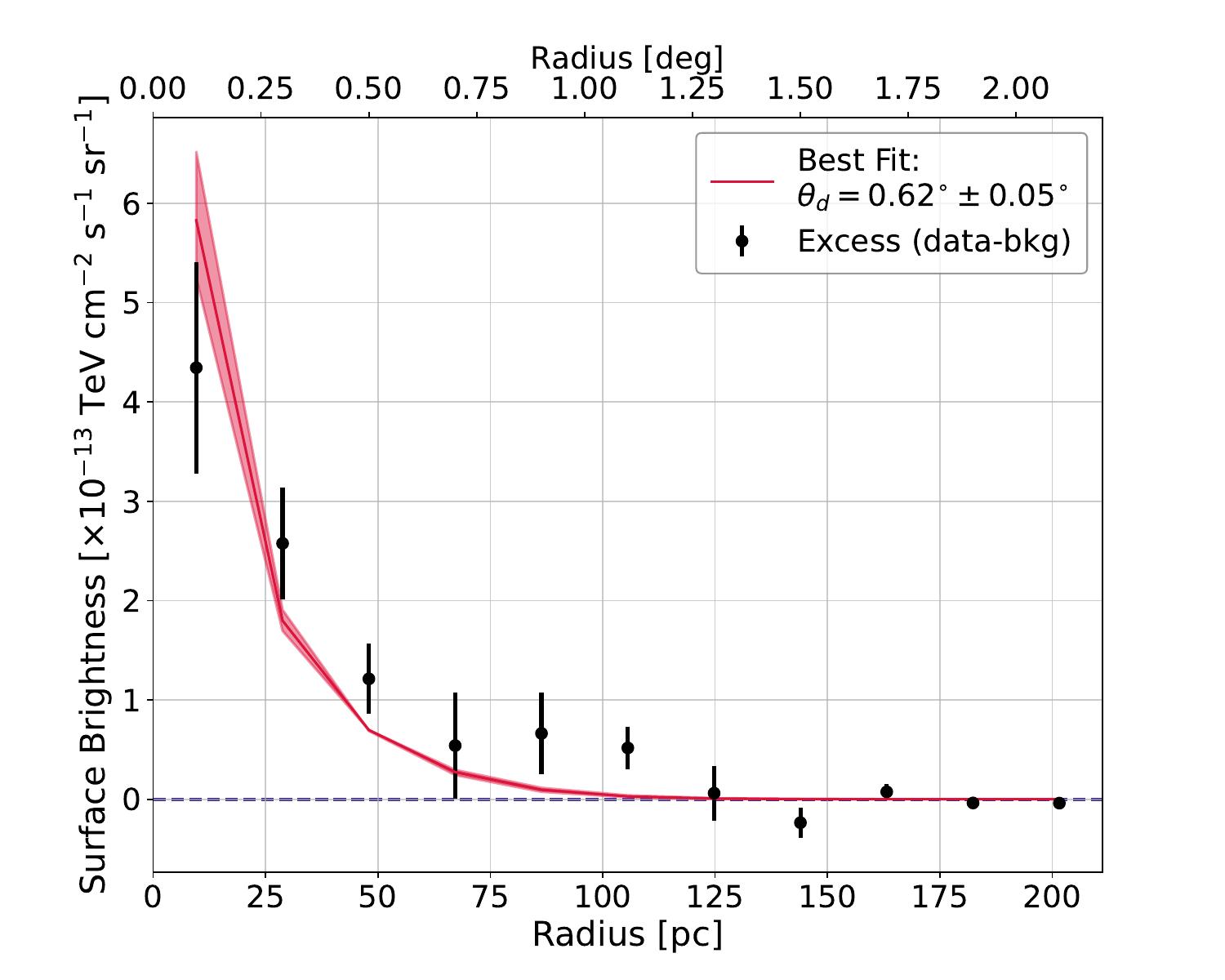}
\end{minipage}
\begin{minipage}[b]{0.33\textwidth}
\centering 
\includegraphics[width=1.0\linewidth]{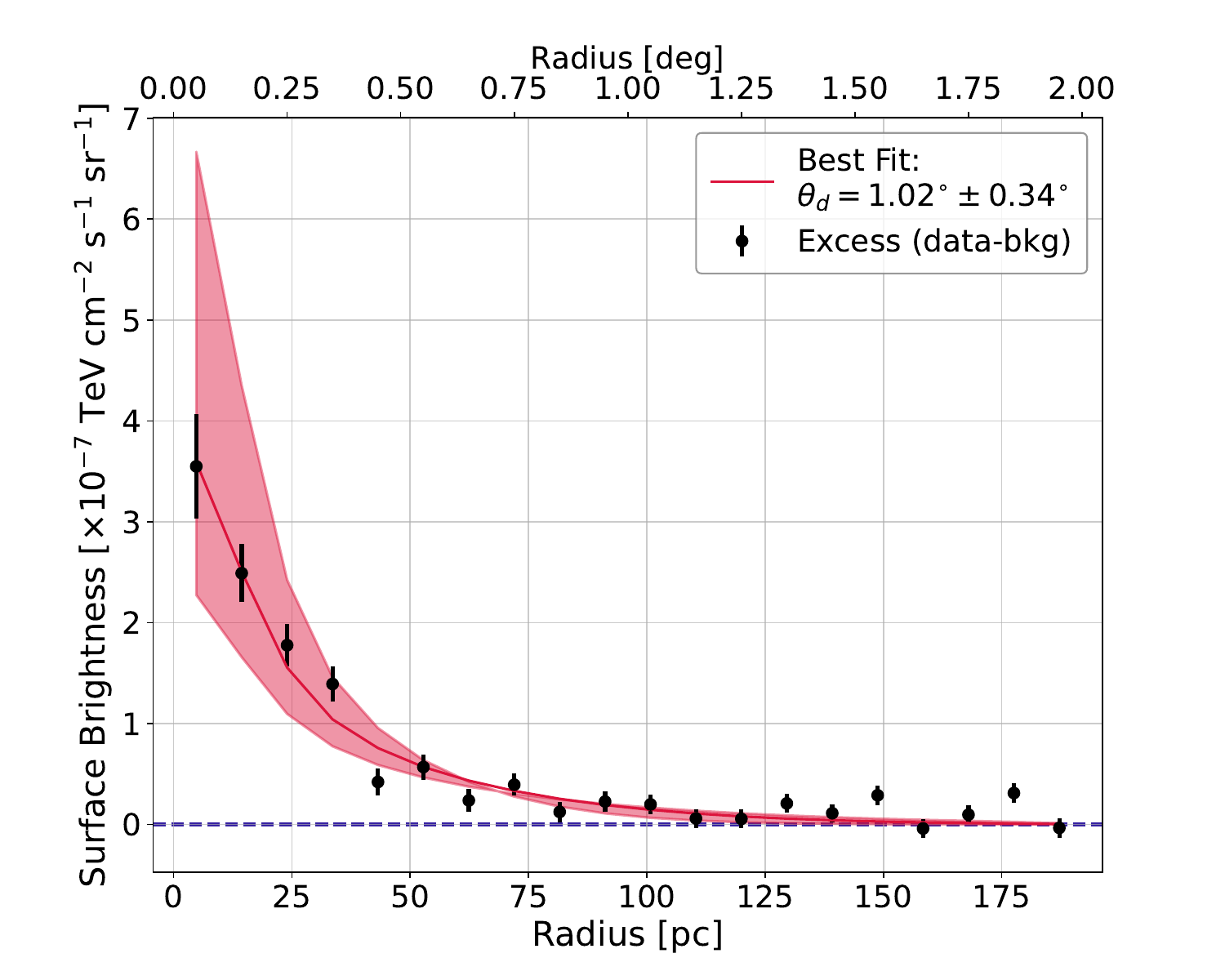}
\end{minipage}
\begin{minipage}[b]{0.31\textwidth}
\centering
    \includegraphics[width=1.0\linewidth]{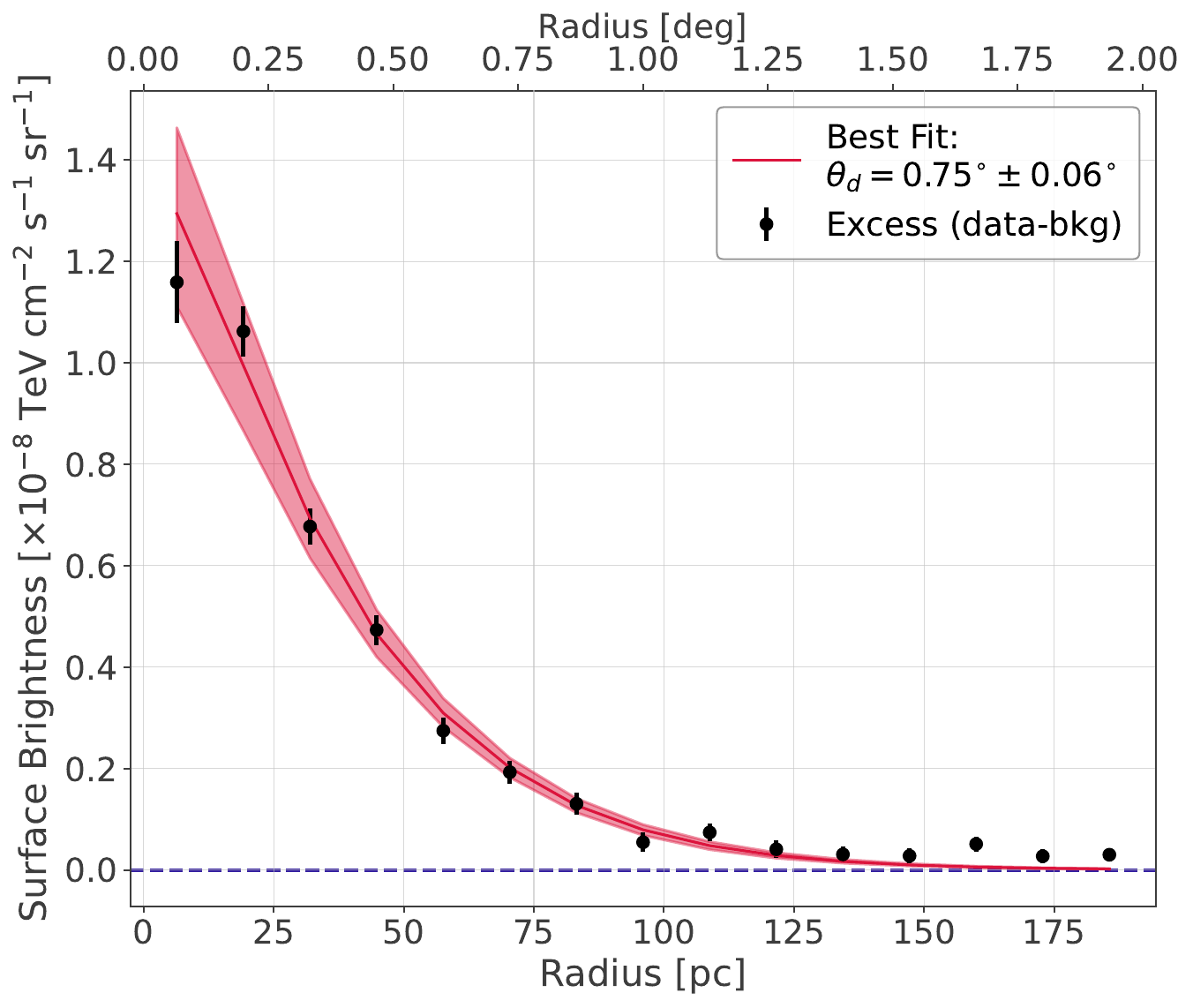}
\end{minipage}
    \caption{\apjrev{Radial surface brightness profiles of HESS J1857+026 for Fermi–LAT data in 10 GeV–2~TeV (left), VERITAS data in 0.3–10~TeV (center), and HAWC data for B=1~$\mu G$ in the 0.67–37~TeV range (right). The profiles are fitted with the diffusion-based surface brightness model given by Equation \ref{equation:ramiro_fluxic}. The theoretical curves are convolved with the instrument PSF during the fitting procedure. The error bars include both statistical uncertainties and the effects of the PSF. For the Fermi–LAT and VERITAS data, the PSF is approximated as a 1D Gaussian with a 1$\sigma$ containment of $0.1^{\circ}$. For HAWC, the PSF is modeled as the sum of two Gaussian functions \citep[see Equation 5 in][]{Albert_2024_crab}}}
\label{fig:radial_brihghtness_profile}
\end{figure*} 
\apjrev{Since cosmic ray diffusion fundamentally depends on particle rigidity rather than energy, we express diffusion measurements in terms of rigidity $R=pc/Ze$ for relativistic particles, where $Ze$ is the particle charge. For electrons, rigidity and energy are nearly equivalent, but we adopt rigidity here to emphasize the general transport scaling.} 

\apjrev{It has been reported that the diffusion coefficient around evolved PWNe, $D_{0,\mathrm{PWN}}$, is $(2$--$30)\times 10^{26}$~cm$^2$~s$^{-1}$ at an electron energy of 1~TeV, which is a factor of $\gtrsim 100$ smaller than the average Galactic diffusion coefficient $D_{0,\mathrm{Gal}}$ \citep{di2020evidences}. This discrepancy may arise from multiple physical effects. One possibility is a change in the rigidity dependence of the diffusion coefficient at high rigidities, such that extrapolations from low-rigidity measurements are no longer valid. Indeed, a softening of the diffusion coefficient scaling index $\delta$ at $\sim 300$~GV has been observed using AMS-02 data \citep{genolini2019cosmic}, supporting this scenario. However, Galactic diffusion measurements primarily probe rigidities between $\sim 2$~GV and $\sim 2$~TV, while TeV halo observations probe much higher rigidities ($\gtrsim 20$~TV), so the extent to which such a change accounts for the full suppression remains unclear. Alternatively, particle transport in the vicinity of sources may differ intrinsically from the average Galactic environment due to locally enhanced magnetic turbulence induced by relativistic particles escaping the PWN. In general, cosmic-ray protons do not undergo significant radiative energy losses and can propagate over large distances before producing observable emission, so observations of hadronic $\gamma$-ray sources primarily probe the cumulative effects of propagation rather than the immediate transport conditions near the source. In contrast, leptonic systems such as PWNe provide a more direct probe of particle transport in the source environment.}

\apjrev{The broadband modeling presented in Section~\ref{sec:yosi} assumes a spatially homogeneous particle distribution and constrains the radiative properties of the system, such as the magnetic field strength and particle spectrum. However, this approach does not capture the spatial transport of particles. In contrast, the extended $\gamma$-ray morphology observed across multiple instruments provides direct information on particle propagation and therefore offers a unique probe of diffusion at these high rigidities around the source.}

\apjrev{In order to investigate the diffusion properties in this regime and to explore the origin of the discrepancy between $D_{\mathrm{PWN}}$ and $D_{\mathrm{Gal}}$, we analyze the radial surface brightness profiles of HESS J1857+026 using Fermi--LAT, VERITAS, and HAWC data across a photon energy range from 10~GeV to 37~TeV, corresponding to an electron rigidity between approximately 1~TV and 200~TV. We therefore adopt a diffusion-based framework to model the spatial distribution of the emission and to constrain the effective diffusion properties of the system. We emphasize that this diffusion modeling is not fully coupled to the radiative modeling in Section~\ref{sec:yosi}; instead, the two approaches provide complementary constraints on the particle population and transport.}

\apjrev{We employ a diffusion-based surface brightness model similar to that used to characterize the Geminga pulsar halo \citep{abeysekara2017extended} to relate the observed $\gamma$-ray spatial profile to the characteristic propagation scale of particles, set by the interplay between diffusion and energy losses. This approach assumes isotropic, continuous injection of electrons and positrons from a point source and models their transport via spatial diffusion. The underlying particle distribution follows the solution of the diffusion equation for continuous injection (e.g., \citealt{Atoyan1995}), while the observed $\gamma$-ray emission is produced through IC scattering on the IRSF, as implemented in \cite{abeysekara2017extended}.}

\apjrev{In this framework, the projected $\gamma$-ray surface brightness profile is described using an analytical approximation to the line-of-sight integral of the diffusing particle distribution, given by}
\begin{equation}
    \frac{dN}{d\Omega} = \frac{1.22}{\pi^{3/2} \theta_d (E_e) [\theta + 0.06\theta_d(E_e)]} \exp \left( -\theta^2 /\theta_d^2 (E_e)\right).
    \label{equation:ramiro_fluxic}
\end{equation}
\apjrev{
\rev{
Here, $N$ is the total photon flux, $\Omega$ is the solid angle, $E$ is the $\gamma$-ray energy, and $\theta$ is the angle from the centroid of the source. $\theta$ is the angular distance from the centroid of the source. The diffusion angle $\theta_d$ represents the characteristic angular scale of the emission, corresponding to the typical propagation distance of particles under diffusion. It is treated as a free parameter in the fit and is related to the physical diffusion radius $r_d$ by}
}
\begin{equation}
    \theta_d = \left(\frac{180}{\pi}\right) \frac{r_d}{d_{src}}
\end{equation}
\apjrev{
\rev{
where $d_{src}$ is the distance to the source. The diffusion coefficient as a function of electron energy $D(E_e)$ is related to the diffusion radius by}
}
\begin{equation}\label{eq:diff_coeff_radius}
    r_d(E_e) = 2 \sqrt{D(E_e) \min \{t_{\rm cool}, t_{\rm inj}}\}
\end{equation}
\apjrev{
\rev{
where the propagation time is the minimum value between the injection and cooling times. The electron energy can then be estimated from the mean $\gamma$-ray photon energy $E$ as described in \cite{abeysekara2017extended,aharonian_very_2004}}
}
\begin{equation} \label{eq:gamma_elec_energy}
    \langle E_{e} \rangle \approx 17 \langle E \rangle^{0.54+0.046\log_{10}\langle E/\text{TeV}\rangle} \text{ TeV}.
\end{equation}
\apjrev{
\rev{
and the diffusion coefficient is formulated as}
}
\begin{equation} \label{eq:diff_energy}
    D(E_e) = D_0 \left(E_e / 10 \ \text{GeV}\right)^{\delta}
\end{equation}
\apjrev{
\rev{
where $\delta$ is the diffusion spectral index and is set to $\frac{1}{3}$ assuming the Kolmogorov turbulence model. $D(E_e)$ is calculated and determines the diffusion radius $r_d$ from Equation~\eqref{eq:diff_coeff_radius}, assuming a magnetic field strength $B$ (which determines the cooling time), an age $t$ (equivalent to the injection time), and distance $d_{src}$. Two values of $B$ are explored for an age $t = 16\,$kyr and distance $d_{src} = 5.5\,$kpc and are discussed below.
}
}
%\apjrev{\st{ In this section, we expand the diffusion model of to consider altogether the new Fermi--LAT, VERITAS, and HAWC results. The diffuse transport of particles from the acceleration site is characterized by the diffusion coefficient $D(E) = D_{0}E^{\delta}$, where $E$ is the particle energy and $\delta$ is the diffusion spectral index. % \rev{(see Section~\ref{sec:diffusion_model})}.}}
%\st{The accelerated electrons and positrons escape the PWN and transport in the ISM via diffusion. The diffusion length of these particles follows Equation, which depends on the minimum value between the injection time $t_{inj}$ (equivalent to the age of the system $t_{\rm age}$) and}
\apjrev{As shown in Equation \ref{eq:diff_coeff_radius}, the diffusion radius is limited by the injection and cooling timescales.}  The cooling time $t_{\rm cool}$ is given by \citep{Moderski_2005},
\begin{equation}\label{eq:cooling_time}
    t_{cool}(\gamma)=\frac{3m_{\rm{e}}c^{2}}{4c\sigma_{T}\gamma}\left(U_{B}+
            \sum_i \frac{U_{\text{rad}, i}}{(1+4\gamma\varepsilon_{0, i})^{3/2}}\right)^{-1},
\end{equation}
where $\gamma$ is the Lorentz factor of electrons and positrons, $m_{\rm{e}}$ is the electron mass, $c$ is the speed of light, $\sigma_{T}$ is the Thompson cross section, and $\varepsilon_{0,i}$ is the average energy of ambient photon components \apjrev{in units of $m_{e}c^{2}$}. $U_{\text{rad},i}$, and $U_{B}$ are the energy densities of the radiation field components and the magnetic field, respectively. We included three components of the interstellar photon field: the CMB $U_{\rm CMB} = 0.26$ eV/cm$^{3}$, infrared $U_{\rm IR} = 0.41$ eV/cm$^{3}$, and optical $U_{\rm opt} = 0.56$ eV/cm$^{3}$ \citep{john_2023_new}. \hrev{There is no radio or X-ray counterpart for HESS~J1857+026, so the magnetic field strength is difficult to determine. For this reason, we} evaluated two magnetic field strengths near the Galactic average: 1~$\mu$G and 5.5~$\mu$G \citep{Unger_2024}. Under the 1~$\mu$G scenario, the cooling time for electrons in $[2,140]$~TV is $[253,20]$~kyr. For the 5.5~$\mu$G case, the range is reduced to $[116,3]$~kyr. This allows us to explore two regimes: (1) when the cooling time exceeds the age of the PWN \apjrev{($\sim$16~kyr)}, in which case the diffusion length is governed by the system's age; and (2) when the cooling time is shorter, becoming the dominant factor limiting the diffusion length.

\apjrev{
\apjrev{We note that Equation~\ref{equation:ramiro_fluxic} provides an approximate solution for the ICS emission profile arising from the diffusion-loss of electron–positron pairs. While the underlying transport framework includes energy losses (e.g., \citealt{Atoyan1995}), the adopted analytical form does not explicitly model these losses in the spatial profile, but instead encapsulates their effects through the characteristic diffusion scale. Subsequent studies incorporating a more realistic spatial model of ICS emission have found consistent results for the diffusion coefficient around two TeV halos Geminga and Monogem \citep[e.g.,][]{Albert_2024}. Therefore, the approximate model should yield results comparable to those obtained with the exact solution.}
}

\apjrev{%\st{Similar to what is done for the 0.3--316\,TeV HAWC data,} 
The diffusion length $r_{d}$ is measured for the Fermi--LAT data between 10\,GeV and 2\,TeV, VERITAS data between 0.3 and 10\,TeV, and HAWC data between 0.67 and 37\,TeV.} \frev{The radial surface brightness profiles for the Fermi--LAT and VERITAS data are fitted using Equation~\ref{equation:ramiro_fluxic} convolved with the instrumental PSF to estimate the diffusion length $r_{d}$. We have utilized a universal PSF of a 1D Gaussian function with 1$\sigma$ containment within 0.1$^\circ$ for both Fermi--LAT and VERITAS instruments. The HAWC fit uses a PSF obtained from the sum of two Gaussian functions \citep[see equation 5 in][]{Albert_2024_crab}.} \frev{The radial profile data and fits are shown in Figure \ref{fig:radial_brihghtness_profile}.} 
%The IC surface brightness radial profile is described with \citep{hawc_2017}: \begin{equation}\label{eq:diffusion_function_1d} f_{\mathrm{IC}}(r,E) = \frac{1}{R_{\mathrm{d}}(E)(r+0.06R_{\mathrm{d}}(E))} \exp \left( -r^{2}/R_{\mathrm{d}}(E)^{2} \right),
%\end{equation}
%where $r$ is the distance from the PWN. 
%The diffusion length of the Fermi and VERITAS energy ranges are obtained by fitting the surface brightness profiles to Equation~\ref{eq:diffusion_function_1d}. 
The best-fit diffusion angle corresponds to a diffusion length of \apjrev{60 $\pm$ 4.8\,pc for Fermi--LAT data and 98 $\pm$ 32\,pc} for VERITAS data. \rev{These values, together with the minimum value between the injection time and cooling time, then determine the diffusion coefficient $D(E)$ following Equation~\ref{eq:diff_coeff_radius}.} %We note that the HAWC data analysis directly measures the electron diffusion length by assuming the diffusion profile described by Equation~\ref{eq:diffusion_function_1d} and is reported in Section~\ref{sec:hawc_analysis} for $B = 1\,\mu$G. 
For $10$~TeV $\gamma$-ray photons detected by HAWC, corresponding to $\sim 65$~TeV electrons, the fitted diffusion length is \apjrev{71 $\pm$ 12\,pc} for a magnetic field strength B = $1\,\mu$G, and is 56 $\pm$ 3.8\,pc for B = $5.5\,\mu$G. All spatial scales are converted from angular extensions, assuming a fixed source distance of 5.5\,kpc.

\begin{table}[ht]
\centering
\begin{tabular}{lcc}
\hline
 & B = 1 $\mu$G & B = 5.5 $\mu$G \\
\hline
$D_{0}$ (Kraichnan)  & 3.12 $\pm$ 0.40 & 5.68 $\pm$ 0.56 \\
$D_{0}$ (Kolmogorov) & 5.55 $\pm$ 0.69 & 10.6 $\pm$ 1.2 \\
\hline
\end{tabular}
\caption{Fitted values with uncertainties of $D_{0}$ for different magnetic field strengths. Units of $D_{0}$ are $10^{27}$ cm$^{2}$ s$^{-1}$.}
\label{tab:diffusion_coefficient}
\end{table}

\begin{figure*}
\begin{minipage}[b]{0.5\textwidth}
\centering 
\includegraphics[width=1.0\linewidth]{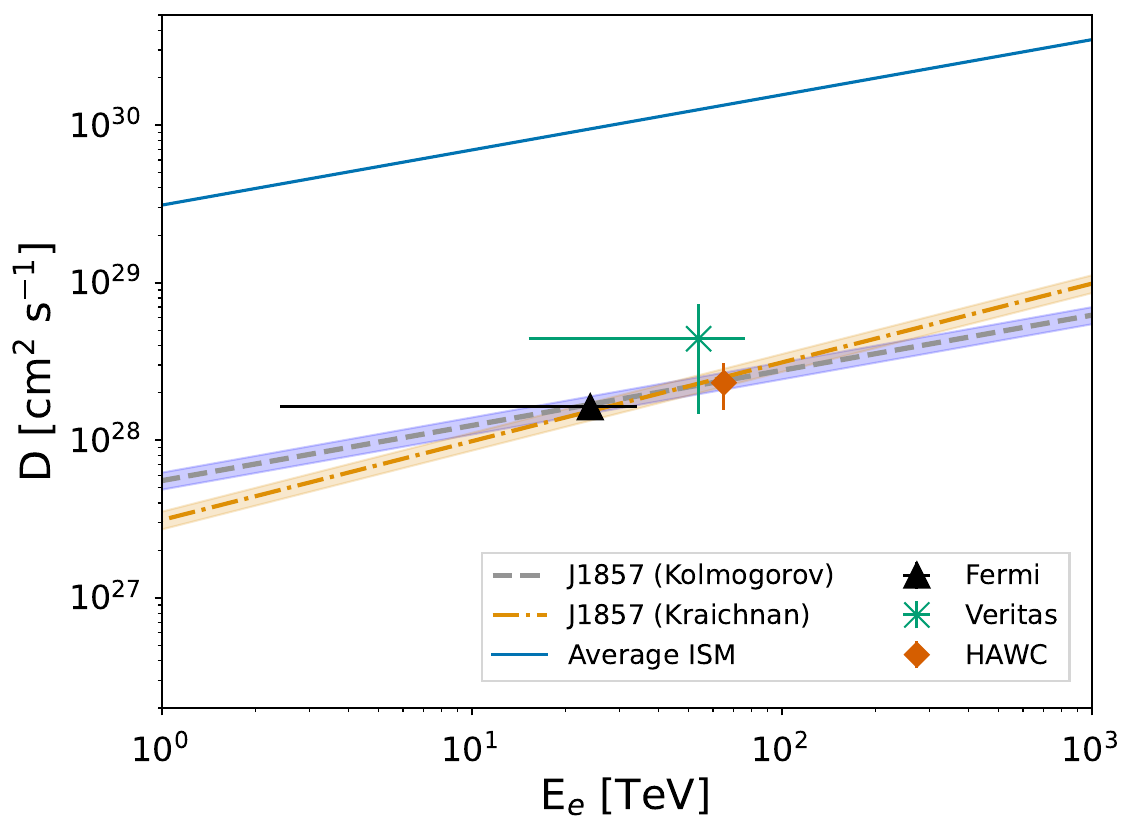}
\end{minipage}
\begin{minipage}[b]{0.5\textwidth}
\centering 
\includegraphics[width=1.0\linewidth]{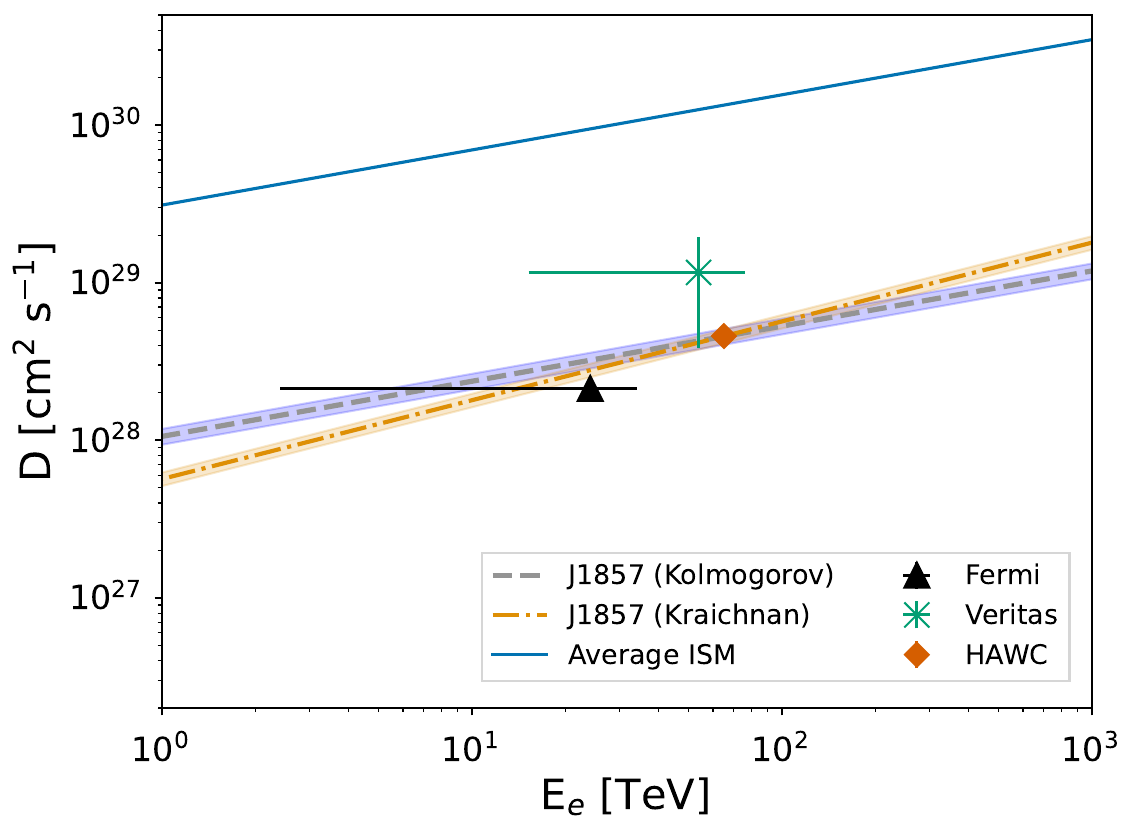}
\end{minipage}
\caption{Diffusion coefficient as a function of electron energy in the environment of HESS~J1857+026. {\it Left:} Assumes a magnetic field of B = 1~$\mu G$. {\it Right:} Assumes a magnetic field of B = 5.5~$\mu G$. The solid line indicates the Galactic average diffusion coefficient following the Kolmogorov ($\delta = 1/3$) regime. The data points are measured from Fermi--LAT, VERITAS, and HAWC radial profiles. The dashed line and the dashed-dotted line are the best-fit diffusion coefficients under the Kolmogorov ($\delta = 1/3$) and Kraichnan ($\delta = 1/2$) diffusion regimes, respectively.}
\label{fig:diffusion_coefficient}
\end{figure*}

The diffusion coefficients are fitted in two diffusion regimes: Kolmogorov ($\delta=1/3$), Kraichnan ($\delta=1/2$). \frev{The fitted value of D$_{0}$ is shown in Table~\ref{tab:diffusion_coefficient}.} The diffusion coefficient in the region of HESS~J1857+026 is compared to the Galactic average ISM value extrapolated for electron energies between 1\,TeV--1\,PeV \citep{Jóhannesson_2019} in Figure~\ref{fig:diffusion_coefficient}. \frev{At the reference electron energy of 50~TeV, the  Galactic diffusion coefficient is $1.2\times10^{30}~\rm{cm}^2/\rm{s}$  assuming Kolmogorov turbulence. At this energy, the diffusion coefficient around the PWN, $D_{\mathrm{PWN}}$, is 
$2.14 \times 10^{28}~\mathrm{cm}^{2}\,\mathrm{s}^{-1}$ for $B = 1\,\mu\mathrm{G}$, and 
$4.08 \times 10^{28}~\mathrm{cm}^{2}\,\mathrm{s}^{-1}$ for $B = 5.5\,\mu\mathrm{G}$, 
under Kolmogorov diffusion. These values correspond to suppression factors 
$D_{\mathrm{PWN}} / D_{\mathrm{ISM}}(50~\rm TeV) = 0.018$ and $0.034$, respectively.} \apjrev{Notably, the measured diffusion coefficients across the Fermi--LAT, VERITAS, 
and HAWC energy ranges are consistent with the standard Kolmogorov 
($\delta = 1/3$) and Kraichnan ($\delta = 1/2$) scaling 
(see Figure~\ref{fig:diffusion_coefficient}), suggesting that the 
suppression is primarily in the normalization $D_0$ rather than in the 
rigidity dependence of the diffusion coefficient. This disfavors a 
scenario in which the discrepancy between $D_{\mathrm{PWN}}$ and 
$D_{\mathrm{Gal}}$ arises from a change in the diffusion scaling index 
$\delta$ at high rigidities, and instead supports locally enhanced 
turbulence near the source as the dominant cause of the suppression.} 

Overall, we find that the diffusion suppression is similar to what is observed for other TeV PWNe \citep {hawc_2017}. A stronger energy dependence in the measured diffusion coefficient is observed for the $B = 1\,\mu$G case (see Figure~\ref{fig:diffusion_coefficient}), which may suggest that the actual magnetic field in the region is closer to this value. For $B = 5.5\,\mu$G, we estimate that electron diffusion above 25\,TeV is limited primarily by the cooling time. This would typically result in a shorter diffusion length for higher-energy electrons, yet an inverse trend is observed in the Fermi--LAT and VERITAS energy ranges relative to HAWC. Within uncertainties, the trend is not statistically significant, mainly limited by the large uncertainties in the VERITAS diffusion size.

It is worth noting that this result assumes particle diffusion is the only transport mechanism into the ISM, which may not fully capture the physical conditions such as potential advection in the region and contamination of the $\gamma$-ray emission due to CRs produced from other sources in the vicinity. %This suggests that there might be contamination of the gamma-ray emission due to CRs from other sources. Other interpretations include extra components of CR transport, such as advection, that exist around the source, thus the diffusion lengths are not implied by Equation. \ref{eq: diffusion length}.

%An example of low- and intermediate-energy analysis is shown in Fig \ref{fig:hess_j1857_Fermi and Fig \ref{fig:hess_j1857_vts}.  Fig \ref{fig:hess_j1857_Fermi shows the $\gamma$-ray sky map and the $\gamma$-ray radial surface brightness with photon energy in $[10,300]$ GeV range using Fermi-LAT data, while Fig \ref{fig:hess_j1857_vts} shows the photon energy in $[0.32,10]$ TeV range using VERITAS data. The radial profile data is fitted with Eq \ref{eq:diffusion_function_1d}, and the best-fit diffusion length is $x$ pc using Fermi-LAT data and 19 pc using VERITAS data, for the distance to the source (PSR J1856+0245) of 5.5 kpc. Assuming a magnetic field of $6\ \mu$G for a middle-age PWN, the cooling time is estimated to be $\sim 13$ kyr, and the derived diffusion coefficient is $D_{\mathrm{PWN}}=2.6 \times 10^{27}\ \mathrm{cm}^{2}/\mathrm{s}$ for photons with energy $>0.3$ TeV.

%\section{Discussion}\label{sec:discussion}
%{\color{red} HESS J1858+020 here?}

\section{Conclusion}\label{sec:conclusion}
We have presented the analysis of high-energy extended emission associated with HESS~J1857+026 using new Fermi--LAT, VERITAS, and HAWC data. The results are complemented with those of HESS \citep{hessgps2018}, MAGIC \citep{magic2014}, and the most recent detection by LHAASO \citep{cao2023first}. A comprehensive view of the high-energy emission depicts a leptonic PWN scenario, supported by the positional coincidence with the energetic pulsar J1856+0245 and \rev{3-sigma indication of} energy-dependent morphology \hrev{in the GeV--TeV band} featuring high-energy $\gamma$-rays concentrating close to the pulsar. A lower-energy $<10\,$GeV spectral component is possible and may be explained by a hadronic origin such as the host or nearby SNR interacting with molecular material in the vicinity \citep{magic2014,petriella2021,guo2023}. \hrev{Molecular gas studies of the region do not indicate a clear correlation between the $\gamma$-ray emission and molecular material. Furthermore,} no SNR shell has been identified to associate or coincide with HESS~J1857+026, challenging \hrev{a hadronic} scenario. \hrev{The low ambient density $n_0 \lesssim 0.13\,$cm$^{-3}$ derived from the models presented in Section~\ref{sec:yosi} also disfavor a hadronic contribution.}

\apjrev{Alternatively, any lower-energy spectral component may arise from a different electron population accelerated within the same PWN} but undergoing different energetic losses, which is consistent with \rev{any} energy-dependent morphology, the \hrev{lack of a PWN X-ray counterpart, and the age of the system, $t \sim [16,21]\,$kyr}. \hrev{Evolved PWNe may have low magnetic field strengths with large electron populations of different cooling times which would generate bright, energy-dependent GeV--TeV emission but faint X-ray emission.} A leptonic PWN origin is therefore favored \apjrev{\citep[see also][]{gong2026}}. Assuming HESS~J1857+026 is the PWN powered by PSR~J1856+026, we characterize the physical and spectral properties of the system using a time-dependent approach that accounts for the basic energetic losses of the particles. %, finding a true age of 21\,kyr and a PWN magnetic field $xx$\,$\mu$G. %The magnetic field estimate provides a basis for a future X-ray detection of the PWN. 
Using the predicted properties from the time-dependent approach combined with the Fermi--LAT, VERITAS, and HAWC surface brightness profiles, we derive the diffusion coefficient $D_{\text{PWN},100\text{TeV}} \sim 10^{28}\,$cm$^{2}$ s$^{-1}$. This value is \apjrev{similar to those found in} other TeV PWNe \citep{hawc_2017} and indicates that diffusion is suppressed relative to the average Galactic ISM in the region around PWNe by $\sim2$ orders of magnitude.

\hrev{A potential neutrino excess is possibly associated, but with an upper limit on the neutrino emission region that is much larger than the $\gamma$-ray region. Future studies investigating the nature of the $\gamma$-ray and neutrino emission can provide further information on the particle accelerator and mechanisms generating the observed emission.}

% \newpage

\begin{acknowledgments}

This publication utilizes data from Galactic ALFA HI (GALFA HI) survey data set obtained with the Arecibo L-band Feed Array (ALFA) on the Arecibo 305m telescope. The Arecibo Observatory is operated by SRI International under a cooperative agreement with the National Science Foundation (AST-1100968), and in alliance with Ana G. Méndez-Universidad Metropolitana, and the Universities Space Research Association. The GALFA HI surveys have been funded by the NSF through grants to Columbia University, the University of Wisconsin, and the University of California.

VERITAS is supported by grants from the U.S. Department of Energy Office of Science, the U.S. National Science Foundation and the Smithsonian Institution, by NSERC in Canada, and by the Helmholtz Association in Germany. This research used resources provided by the Open Science Grid, which is supported by the National Science Foundation and the U.S. Department of Energy's Office of Science, and resources of the National Energy Research Scientific Computing Center (NERSC), a U.S. Department of Energy Office of Science User Facility operated under Contract No. DE-AC02-05CH11231. We acknowledge the excellent work of the technical support staff at the Fred Lawrence Whipple Observatory and at the collaborating institutions in the construction and operation of VERITAS. 
%R.S. thanks NSF for support under NSF grants PHY-XXX at UCLA and PHY-2110497 at Barnard College. 
The authors thank K. Mori for constructive comments.
\end{acknowledgments}

\begin{acknowledgments}
HAWC acknowledges the support from: the US National Science Foundation (NSF); the US Department of Energy Office of High-Energy Physics; the Laboratory Directed Research and Development (LDRD) program of Los Alamos National Laboratory; Consejo Nacional de Ciencia y Tecnolog\'ia (CONACyT), M\'exico, grants 271051, 232656, 260378, 179588, 254964, 258865, 243290, 132197, A1-S-46288, A1-S-22784, c\'atedras 873, 1563, 341, 323, Red HAWC, M\'exico; DGAPA-UNAM grants IG101320, IN111716-3, IN111419, IA102019, IN110621, IN110521; VIEP-BUAP; PIFI 2012, 2013, PROFOCIE 2014, 2015; the University of Wisconsin Alumni Research Foundation; the Institute of Geophysics, Planetary Physics, and Signatures at Los Alamos National Laboratory; Polish Science Centre grant, DEC-2017/27/B/ST9/02272; Coordinaci\'on de la Investigaci\'on Cient\'ifica de la Universidad Michoacana; Royal Society - Newton Advanced Fellowship 180385; Generalitat Valenciana, grant CIDEGENT/2018/034; The Program Management Unit for Human Resources \& Institutional Development, Research and Innovation, NXPO (grant number B16F630069); Coordinaci\'on General Acad\'emica e Innovaci\'on (CGAI-UdeG), PRODEP-SEP UDG-CA-499; Institute of Cosmic Ray Research (ICRR), University of Tokyo, H.F. acknowledges support by NASA under award number 80GSFC21M0002. We also acknowledge the significant contributions over many years of Stefan Westerhoff, Gaurang Yodh and Arnulfo Zepeda Dominguez, all deceased members of the HAWC collaboration. Thanks to Scott Delay, Luciano D\'iaz and Eduardo Murrieta for technical support.
\end{acknowledgments}

\begin{acknowledgments}
The Fermi--LAT Collaboration acknowledges generous ongoing support
from a number of agencies and institutes that have supported both the
development and the operation of the LAT as well as scientific data analysis.
These include the National Aeronautics and Space Administration and the
Department of Energy in the United States, the Commissariat \`a l'Energie Atomique
and the Centre National de la Recherche Scientifique / Institut National de Physique
Nucl\'eaire et de Physique des Particules in France, the Agenzia Spaziale Italiana
and the Istituto Nazionale di Fisica Nucleare in Italy, the Ministry of Education,
Culture, Sports, Science and Technology (MEXT), High Energy Accelerator Research
Organization (KEK) and Japan Aerospace Exploration Agency (JAXA) in Japan, and
the K.~A.~Wallenberg Foundation, the Swedish Research Council and the
Swedish National Space Board in Sweden. Additional support for science analysis during the operations phase is gratefully
acknowledged from the Istituto Nazionale di Astrofisica in Italy and the Centre
National d'\'Etudes Spatiales in France. This work performed in part under DOE
Contract DE-AC02-76SF00515.
\end{acknowledgments}

% \clearpage
\software{FermiPy \citep[v.1.2.0][]{fermipy2017}, Fermitools: Fermi Science Tools \citep[v2.2.11][]{fermitools2019}, Eventdisplay \citep{maier2017eventdisplay}\citep{Maier_Eventdisplay_An_Analysis_2024}, Gammapy \citep{gammapy:2023,gammapy_zenodo_v1p3}, NAIMA \citep{naima}}

\appendix
\restartappendixnumbering
\section{Supplemental HAWC Analysis Results}
\hrev{Table \ref{tab:hawc-res-5ug} shows the HAWC best-fit results assuming a magnetic field value of 5 $\mu$G for the diffusion modeling of HESS J1857+026 (see Section~\ref{sec:hawc_analysis} for details). In this case, the diffusion radius is determined mainly by the  electron cooling time, which is on the order of a few kyr (see Section \ref{sec:pwn_physics}).}

\begin{table}
\centering
\begin{tabular}{l c}
\hline
\hline
Parameter  & Best-fit value \\
%& (Diffusion Model) \\
\hline
& HAWC J1854+0120 \\
\hline
         $\sigma$ & $0.^{\circ}69\pm0.06_{\text{stat}} \pm 0.25_{\text{syst}}$  \\
         $K$ & $16.6(_{-2.3}^{+2.6})_{\text{stat}}(^{+12}_{-9})_{\text{syst}}$\\
         $\Gamma$ & $2.73\pm0.04_{\text{stat}} \pm 0.09$ \\
\hline
& HAWC J1857+0200 (HESS J1858+020) \\
         \hline
        $\sigma$ & $0.^{\circ}20\pm0.01_{\text{stat}} \pm 0.03_{\text{syst}}$ \\
         $K$ & $9.2\pm0.6_{\text{stat}} \pm 1.7_{\text{syst}} $ \\
         $\Gamma$ & $2.45\pm0.03_{\text{stat}} \pm 0.08_{\text{syst}}$ \\
\hline
& HAWC J1857+0247 (HESS J1857+026) \\
    \hline
         $\sigma$ & $0.^{\circ}60\pm0.04_{\text{stat}}\pm0.14_{\text{sys}}$ \\
         $K$ & $52 \pm 5_{\text{stat}}(^{+32}_{-25})_{\text{syst}}$ \\
         $\Gamma$ & $2.20\pm0.05_{\text{stat}}\pm 0.32_{\text{syst}}$\\
         $E_c$ & $12(_{-1.0}^{+1.1})_{\text{stat}}(^{+7.7}_{-5.9})_{\text{syst}}$ \\
         \hline
     &   HAWC J1858+0344 \\
\hline
         $\sigma$ & $0.^{\circ}59\pm0.05_{\text{stat}}\pm 0.15_{\text{syst}}$ \\
         $K$ & $13.5(_{-1.7}^{+2.0})_{\text{stat}}(^{+6.9}_{-5.5})_{\text{syst}}$\\
         $\Gamma$ & $2.65\pm0.03_{\text{stat}} \pm 0.08_{\text{syst}} $\\
        \hline
        $K_{\text{GDE}}$ &  $1.55\pm0.18_{\text{stat}}\pm 0.75_{\text{syst}}$ \\
\hline
\hline
\end{tabular}
\caption{HAWC best-fit results in the 0.7--37\,TeV energy range for the diffusion model of HESS~J1857+026 assuming $B=$5\,$\mu$G. The normalization flux values $K$ have units $\times 10^{-15} \ \rm{TeV}^{-1}\,\rm{cm}^{-2}\,\rm{s}^{-1}$. $E_c$ is in TeV. $\sigma$ corresponds to the Gaussian width.
}
\label{tab:hawc-res-5ug}
\end{table}

\end{document}